\documentclass[twocolumn,showpacs,english,superscriptaddress,
preprintnumbers,amsmath,amssymb,floatfix]{revtex4-1}

\usepackage[T1]{fontenc}
\usepackage[latin9]{inputenc}
\usepackage{dcolumn}
\usepackage{bm}
\usepackage{graphicx}
\usepackage{color}
\usepackage{esint}
\usepackage{babel}
\usepackage{amsfonts}
\usepackage{slashed}
\usepackage{bm}
\usepackage{amsmath}
\usepackage{epsfig}
\usepackage{enumerate}
\def\be{\begin{equation}}
\def\ee{\end{equation}}

\begin{document}

\title{Josephson effect in multiterminal topological junctions}

\author{A.~Zazunov}
\affiliation{Institut f\"ur Theoretische Physik, Heinrich-Heine-Universit\"at, D-40225  D\"usseldorf, Germany}

\author{R.~Egger}
\affiliation{Institut f\"ur Theoretische Physik, Heinrich-Heine-Universit\"at, D-40225  D\"usseldorf, Germany}

\author{M.~Alvarado}
\affiliation{Departamento de F{\'i}sica Te{\'o}rica de la Materia Condensada C-V,
Condensed Matter Physics Center (IFIMAC) and Instituto Nicol\'as  Cabrera,
 Universidad Aut{\'o}noma de Madrid, E-28049 Madrid, Spain}

\author{A.~Levy Yeyati}
\affiliation{Departamento de F{\'i}sica Te{\'o}rica de la Materia Condensada C-V,
Condensed Matter Physics Center (IFIMAC) and Instituto Nicol\'as  Cabrera,
 Universidad Aut{\'o}noma de Madrid, E-28049 Madrid, Spain}

\date{\today}

\begin{abstract}
We study the Josephson effect in a trijunction formed by two topological superconductor (TS) wires and a conventional $s$-wave superconductor.
Using a boundary Green's function formalism, analytical results for the current-phase relation are obtained in various limiting cases by 
modeling the  TS wires via the low-energy limit of a Kitaev chain.  We show that Josephson transport critically depends on the spin canting angle $\theta$ between
the boundary spin polarizations of the TS wires, which in turn suggests that the spin structure of Majorana states can be accessed
through supercurrent measurements. We also extend the boundary Green's function approach to a more microscopic spinful wire model
and thereby compute the dependence of $\theta$ on experimentally accessible parameters such as the Zeeman field and/or the
chemical potential.  Furthermore, we show that the equilibrium current-phase relation between both TS wires exhibits a robust $4\pi$-periodicity
since the conventional superconducting lead effectively locks the fermion parity of the trijunction. 
\end{abstract}


\maketitle

\section{Introduction}\label{sec1}

The possibility to induce topological superconductivity in semiconducting nanowires
with strong spin-orbit coupling (e.g., InAs or InSb) by proximity coupling to a
conventional superconductor has paved the way to exciting novel developments
\cite{Kitaev2001,Alicea2012,Leijnse2012,Beenakker2013,Aasen2016,Mourik2012,Das2012,
Albrecht2016,Deng2016,Guel2017,Albrecht2017,Zhang2017,Suominen2017}. 
In particular,  presently available devices with a hard superconducting gap \cite{Albrecht2016,Deng2016,Guel2017,Albrecht2017} 
and disorder levels close to the ballistic limit 
\cite{Zhang2017,Suominen2017} exhibit characteristic transport signatures  as
expected for Majorana bound states (MBSs). The latter are robust states localized at the wire ends
within the topological phase. 
Similar experiments have also been reported for other platforms, see for instance Refs.~\cite{Yazdani2014,Franke2015,Sun2016,Feldman2017,Deacon2017}, and 
many of our results below apply to those implementations as well.
It stands to reason that the currently developed technologies will give access to
quantum transport studies in a wide variety of hybrid structures 
involving topological superconductor (TS) wires, including multiterminal geometries \cite{Zhang2017,Suominen2017,Gazi2017}. 
Such devices may also be of interest for topological quantum information processing applications \cite{Alicea2012,Leijnse2012,Beenakker2013}.

This situation claims for a flexible theoretical 
framework going beyond idealized models but simple enough to account in an (almost) analytical way
 both for subgap (Andreev and/or Majorana) bound states as well as for above-gap continuum states. 
In a recent work \cite{Zazunov2016}, an effective low-energy theory has been formulated in order to analyze nonequilibrium quantum transport in junctions of 
TS wires and/or conventional (normal or superconducting) terminals. 
Based on a tunnel Hamiltonian description valid for arbitrary junction transparency, the boundary Green's function (bGF) of the respective terminal here represents a
 central ingredient, see also Ref.~\cite{Peng2017}. 
It is well known that related approaches can successfully describe a wide variety of topologically trivial systems \cite{Cuevas1996,Nazarov2009}.
Analytical results for the (nonlinear) conductance, noise correlations, and for the supercurrent-phase relation in topological two-terminal 
 hybrid junctions have been reported in Ref.~\cite{Zazunov2016}. 
 Moreover, a three-terminal junction with two normal leads and one TS wire has been studied within this framework  \cite{Jonckheere2017}.
  
\begin{figure}[t]
\centering
\includegraphics[width=0.45\textwidth]{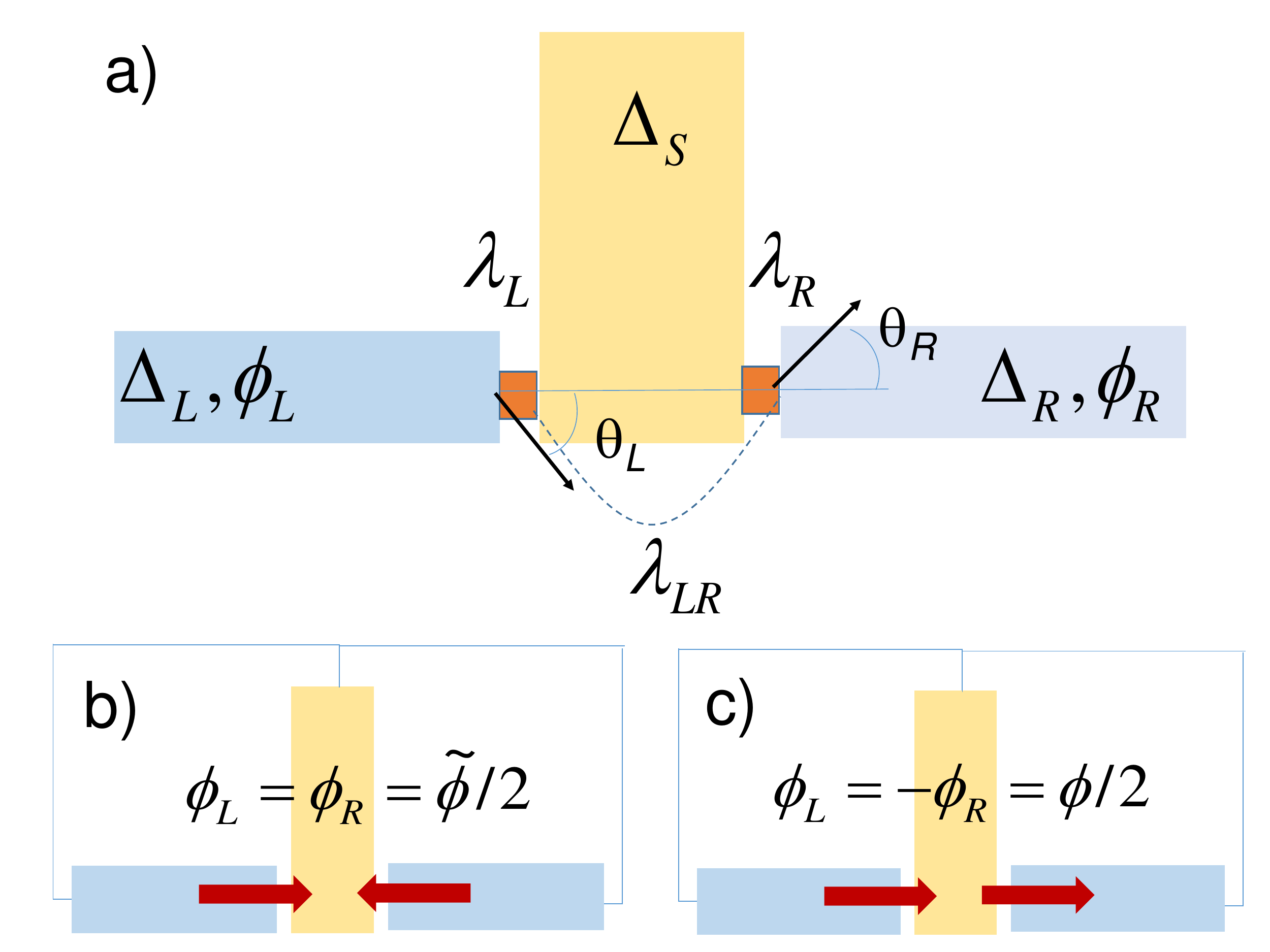}
\caption{\label{f1}
(a) \textit{Schematic setup:} Phase-biased TS-S-TS trijunction formed from  
 two topological superconductor (TS) wires  and one conventional superconductor (S).
 The TS wires (blue) have the respective superconductor order parameter
 $\Delta_{L/R}e^{i\phi_{L/R}}$, while the S terminal has the gap $\Delta_S$
 and phase $\phi_S=0$.  The S-TS tunnel couplings  
 are denoted by $\lambda_{L/R}$, and the direct TS-TS coupling is  
$\lambda_{LR}$.  The spin quantization axis in the respective TS wire is tilted by an
angle  $\theta_{L/R}$ relative to the corresponding axis in the S lead.
(b) \textit{Parallel configuration:} One chooses  $\phi_L=\phi_R=\tilde \phi/2$ such that $\phi=0$, see Eq.~\eqref{phasedef}.
Red arrows indicate supercurrents $I_{L/R}$ in TS wires.
(c) \textit{Serial configuration:} Here $\phi_L=-\phi_R=\phi/2$ with $\tilde \phi=0$, see Eq.~\eqref{phasedef}.}
\end{figure}

The present work aims at providing a concise theoretical description
of quantum transport in multiterminal junctions containing TS wires as well as 
conventional $s$-wave superconducting leads (hereafter referred to as S leads).
We focus on the TS-S-TS setup sketched in Fig.~\ref{f1}(a), 
where we analyze the equilibrium supercurrent-phase relation in a
phase-biased device.  In addition, we go beyond previously established theories by
deriving and subsequently employing the bGF  for  
the spinful semiconductor nanowire model proposed in Refs.~\cite{Lutchyn2010,Oreg2010}.
Compared to the low-energy bGF for effectively spinless Kitaev chains used in Refs.~\cite{Zazunov2016,Jonckheere2017}, this 
more microscopic bGF allows one to directly calculate the spin-dependent properties of a TS wire.  While our results for the spinful 
 bGF apply for arbitrary multiterminal junctions, spin-dependent properties
turn out to be of special importance for the TS-S-TS setup in Fig.~\ref{f1}.
Since we study equilibrium properties, we utilize the Matsubara representation throughout this paper.
By instead using a Keldysh version of the theory \cite{Zazunov2016}, 
nonequilibrium features in topological multiterminal junctions, e.g., phenomena involving multiple Andreev reflection (MAR) processes,
may also be studied. However, we leave this to future work. 

The Josephson effect in  hybrid S-TS multiterminal junctions constitutes an interesting fundamental issue. Since Majorana states 
have a well-defined spin polarization axis \cite{Flensberg2011,Sticlet2012,Prada2012,Rainis2013,Jiang2013,
He2014,Prada2017,Hoffman2017}, Josephson transport as well as MAR phenomena are
strongly suppressed in two-terminal S-TS junctions 
 \cite{Zazunov2012,Peng2015,Ioselevich2016,Sharma2016,Setiawan2017a,Setiawan2017b}. 
This supercurrent blockade reflects the conflicting pairing symmetries in both leads. However, the
blockade could be lifted by forming a junction between an S lead and two (or more) TS leads. In this case,  a spin-singlet Cooper
pair in S can be split by injecting two opposite-spin electrons into different TS leads (or
the reverse process).   Clearly, such processes, and thus the supercurrent, will then be
sensitive to the relative angle $\theta=\theta_L-\theta_R$ between the spin polarization axes of both
TS wires, see Fig.~\ref{f1}(a).  If the spins in the TS wires are oriented along the same direction ($\theta=0$), the supercurrent blockade will persist,
while a sizable supercurrent can flow for antiparallel alignment ($\theta=\pi$).
Since each TS wire effectively acts as a spin filter, it is not possible to generate spin entanglement in this way. However, 
Cooper pairs can be split with high efficiency through such processes \cite{Nilsson2008}, and
supercurrent measurements  in the trijunction setup of Fig.~\ref{f1}(a) can therefore probe the spin structure associated with MBSs.
We mention in passing that previous work has considered networks of TS wires 
\cite{Alicea2011},  where TS-TS-TS trijunctions represent the basic unit~\cite{Jiang2011}.  
Topological aspects may also become important in multiterminal junctions of conventional superconductors \cite{Riwar2016}.
However, to our knowledge, multiterminal geometries containing both TS and S leads have not been discussed before.

We thus consider a conventional $s$-wave superconductor (S) and two topological class-$D$ superconductor (TS) wires,
 which are tunnel-coupled in a trijunction geometry, see Fig.~\ref{f1}(a). Here,
 $\Delta_S$ and $\Delta_{j=L/R}$ are real-valued superconducting gaps, where 
gauge invariance allows us to put the phase in the S lead to
$\phi_S=0$ without loss of generality.  Taking into account current conservation, 
\begin{equation}\label{currentcons}
I_L+I_R+I_S=0,
\end{equation}
with the convention that individual supercurrents flowing through each branch ($L/R/S$) are oriented towards the junction,
it is useful to introduce the linear combinations 
\begin{equation}\label{phasedef}
\phi= \phi_L-\phi_R,\quad \tilde \phi= \phi_L+\phi_R.
\end{equation}

For $\phi=0$, we have the ``parallel configuration'' shown in Fig.~\ref{f1}(b), where $I_L=I_R$ implies that no supercurrent flows between the TS leads. 
We then effectively have a Josephson junction between the $s$-wave superconductor and
a two-channel $p$-wave superconductor. The phase difference across this effective junction is given by $\tilde \phi/2$.

 For $\tilde \phi=0$,  we instead encounter the ``serial configuration'' in Fig.~\ref{f1}(c), where the supercurrent $I_R=-I_L$ 
 flows between both TS wires but $I_S=0$. The S lead then acts as currentless probe electrode (with gapped spectrum) coupled to an effective TS-TS junction.
 Remarkably, as we show in Sec.~\ref{sec2c} below, since the S lead effectively measures the fermion parity of the trijunction, the $4\pi$-periodic 
 current-phase relation of a TS-TS junction \cite{Kitaev2001,Alicea2012,Leijnse2012,Beenakker2013} is stabilized for all $\theta\ne 0$ 
 and thereby becomes completely robust.
 It is worth noting that this effect cannot be captured by 
an approach neglecting above-gap quasiparticles.
  This observation suggests a practical way to experimentally observe the topological Josephson current-phase relation (CPR) in a ground state
 of fixed fermion parity.
 
 In what follows, we always take $\Delta_L = -\Delta_R = \Delta$ as expected for a long TS wire interrupted by a tunnel junction \cite{Zazunov2016}
and mostly assume symmetric S-TS couplings, $\lambda_L = \lambda_R= \lambda$, keeping the direct TS-TS coupling $\lambda_{LR}$ arbitrary. 
The crucial parameter controlling the Josephson effect is then given by the spin canting angle $\theta$, see Fig.~\ref{f1}.

Our theoretical description of TS-S-TS trijunctions below proceeds along two interrelated avenues. In Sec.~\ref{sec2},
we use the bGF for TS wires derived in Ref.~\cite{Zazunov2016}, see Eq.~\eqref{bGFK} below, which in turn represents the 
low-energy limit of the celebrated Kitaev chain \cite{Kitaev2001}. 
In this case, the spin structure is introduced phenomenologically through the angles $\theta_{L/R}$ in Fig.~\ref{f1}(a). 
The advantage of this approach is that its simplicity allows for analytical results.
 As a second approach, in Sec.~\ref{sec3}, we derive and employ the bGF for
the spinful TS nanowire model in Refs.~\cite{Lutchyn2010,Oreg2010}. 
This model explicitly takes into account proximity-induced pairing,  spin-orbit coupling, and the magnetic Zeeman field. 
 Details of our derivation of the bGF can be found in the Appendix. Using this bGF, the MBS spin structure no longer needs
to be inserted in a phenomenological manner but emerges naturally. However, this advantage comes at the prize of increased complexity, necessitating  a 
numerical analysis.  We also study the CPR across the topological transition in the TS wires using this approach. 
Finally, we conclude in Sec.~\ref{sec4}.    Throughout, we use units with $e=\hbar=k_B=1$.

\section{Low-energy approach}\label{sec2}

\subsection{Model}\label{sec2a}

In this section, each TS wire is modeled by the bGF in Eq.~\eqref{bGFK} below, which describes 
the  low-energy continuum limit of an effectively spinless 
Kitaev chain deep in the topological phase, see Ref.~\cite{Zazunov2016} for a detailed derivation.
The spin angles $\theta_{L/R}$ are then taken into account phenomenologically in the tunnel Hamiltonian, see Eq.~\eqref{Ht} below.
Starting from the boundary fermion operator $c_j$ in TS lead $j=L/R$, we define
 the Nambu spinor $\Phi_j = \left(c^{}_j, c^\dagger_j\right)^T$. Using the 
time-ordering operator ${\cal T}$, the imaginary-time bGF of an uncoupled TS lead is
\begin{eqnarray}\nonumber
G_{j=L/R}(\tau) &=& - \langle {\cal T} \Phi_j^{} (\tau) \Phi^\dagger_j(0) \rangle^{}_0 =T \sum_\omega e^{-i \omega \tau} G_j(\omega),\\
\label{bGFK}
G_j(\omega) &=& \frac{1}{i \omega}\left( \sqrt{\omega^2 + \Delta_j^2} \, \sigma_0 + \Delta_j \sigma_x \right) ,
\end{eqnarray}
where Pauli matrices $\sigma_{x,y,z}$ and the identity $\sigma_0$ act in Nambu space. All frequency summations
run over fermionic Matsubara frequencies, $\omega=2\pi (n+1/2) T$, with integer $n$ and temperature $T$.
Similarly,  we have boundary fermion operators $\psi_\sigma$ for spin $\sigma$ in the S lead,
which are combined to form the Nambu spinor  $\Psi = \left(\psi_\uparrow^{}, \psi^\dagger_\downarrow \right)^T$.
The respective bGF is then given by \cite{Nazarov2009}
\begin{eqnarray}\nonumber
G_S(\tau) &=& - \langle {\cal T} \Psi(\tau) \Psi^\dagger(0) \rangle^{}_0 = T \sum_\omega e^{-i \omega \tau} G_S(\omega),\\
\label{bGFS}
G_S(\omega) &=& - \frac{i \omega \sigma_0 + \Delta_S \sigma_x}{\sqrt{\omega^2 + \Delta_S^2}}.
\end{eqnarray}

In the above formulation, we have used a gauge where the superconducting phases $\phi_{L/R}$, see Fig.~\ref{f1}(a),
 only appear in the tunnel
Hamiltonian connecting the decoupled TS and S leads  in order to
form a trijunction,
\begin{eqnarray}\label{Ht}
H_t  &=&  \sum_{j = L/R} \lambda_j e^{i \phi_j/2}\left( \alpha_j \psi^\dagger_\uparrow +  \beta_j \psi^\dagger_\downarrow \right) c_j  \\
\nonumber &+& 
\lambda_{LR} e^{-i \phi/2} c^\dagger_L c^{}_R + {\rm h.c.},
\end{eqnarray}
where the real-valued S-TS tunnel amplitudes
 $\lambda_{L/R}\ge 0$ include density-of-state factors, and the prefactors 
$\alpha_{j} = e^{i \chi_j} \cos ( \theta_j/2)$ and $\beta_j = e^{-i \chi_j} \sin( \theta_j/2)$
 determine the weight of each spin component through the angles $\theta_{L/R}$.
 In addition, there are angles $\chi_{L/R}$ which, however, are not essential as shown below.
Furthermore, $\lambda_{LR}$ in Eq.~\eqref{Ht} describes a direct coupling between both TS leads.
With the above conventions, all tunnel couplings are dimensionless quantities. 
In Nambu notation, Eq.~\eqref{Ht} is  given by
\begin{eqnarray}\label{Ht2}
H_t  &=& \frac12  \Phi^\dagger_L W_{LR}\Phi_R^{} +  \sum_{j=L/R} \Psi^\dagger_{} W_j \Phi^{}_j  + {\rm h.c.},\\
\nonumber
W_{LR} &=& \lambda_{LR} e^{-i \sigma_z \phi/2} \sigma_z, \quad 
W_{j} = 
\lambda_j e^{i \sigma_z \phi_j/2} {\rm diag}( \alpha_j , -\beta_j^\ast).
\end{eqnarray}
Let us now take the global spin quantization axis along the spin direction of, say, the left ($j=L$) TS lead \cite{foot1}.
We then observe that SU(2) spin rotation invariance in the S lead implies that
we can set $\alpha_L=1$ and $\beta_L=0$ by switching to new fermion fields for the S terminal,
$\psi_\uparrow' = \alpha_L^\ast \psi_\uparrow + \beta_L^\ast \psi_\downarrow$ and
$\psi_\downarrow' = -\beta_L \psi_\uparrow + \alpha_L \psi_\downarrow$.
Another simplification follows by noting that $\chi_R$ can be gauged away up to a renormalization of the phase $\phi_L$.
We therefore put $\alpha_R=\cos(\theta/2)$ and $\beta_R=-\sin(\theta/2)$ with the angle $\theta=\theta_L-\theta_R$ between the TS spin polarization directions.
Note that the direct TS-TS coupling $\lambda_{LR}$ will in principle 
depend on $\theta$.  In particular, if only spin-conserving tunneling is 
possible, we expect $\lambda_{LR}=0$ for $\theta=\pi$. 

The Josephson currents $I_{L/R}$, and thus also $I_S$ from Eq.~\eqref{currentcons}, can then be expressed as \cite{Zazunov2016}  
\begin{equation}\label{Ij}
I_j = - i T\ {\rm tr}\sum_\omega  \sigma_z \left[ \Sigma(\omega) {\cal G}(\omega) \right]_{jj} .
\end{equation}
The full bGF,  ${\cal G}(\omega)$, follows from the uncoupled bGFs in Eq.~\eqref{bGFK} as a solution of the Dyson equation,
\begin{equation}\label{dyson} 
{\cal G}^{-1}(\omega) = G^{-1}(\omega) - \Sigma(\omega),
\end{equation}
where the matrix $G ={\rm diag}(G_L, G_R)$ acts in TS lead space and
the trace operation in Eq.~\eqref{Ij} is over Nambu space.
The  self-energy $\Sigma(\omega)$ captures the effects of integrating out the S lead. Like ${\cal G}(\omega)$, this self-energy is  a $4 \times 4$ matrix
in lead-Nambu space.
Using the particle-hole symmetry relation $\Phi_j^\dagger = \Phi_j^T \sigma_x$, some algebra gives the self-energy matrix elements in lead space
in terms of the Nambu matrices
\begin{eqnarray}\nonumber
&& \Sigma_{LL/RR}(\omega)  =  - \frac{i \omega \lambda_{L/R}^2}{ \sqrt{\omega^2 + \Delta_S^2}} \, \sigma_0 ,
\\ \label{selfen}
&&\Sigma_{LR}(\omega)  =\Sigma^{\dagger}_{RL}(-\omega)=W_{LR}-  \frac {\lambda_L \lambda_R}{\sqrt{\omega^2 + \Delta_S^2}}\times\\
&& \times
\left( \begin{array}{cc} i \omega \cos(\theta/2)  e^{-i \phi/2} &  \sin(\theta/2) \Delta_S e^{-i \tilde \phi/2} \\
-\sin(\theta/2) \Delta_S e^{i \tilde \phi/2} & i \omega \cos(\theta/2) e^{i \phi/2} \end{array} \right), \nonumber
\end{eqnarray}
with $\phi$ and $\tilde \phi$ in Eq.~\eqref{phasedef} and $W_{LR}$ in Eq.~\eqref{Ht2}.
Notably, the Cooper pairing term mediated between both TS wires through lead S, which corresponds to the off-diagonal Nambu component in 
Eq.~\eqref{selfen}, exhibits a pronounced $\theta$-dependence through the factor $\sin (\theta/2)$.
In particular, the induced pairing is maximal for opposite spin polarization of the TS leads ($\theta=\pi$) but
 vanishes for collinear polarization ($\theta=0$). 

A numerical calculation of the CPR in Eq.~\eqref{Ij} poses no serious challenges, but fortunately
many limiting cases of interest can also be treated analytically.  
In the remainder of this section, we first discuss the parallel configuration with $\phi=0$, see Fig.~\ref{f1}(b) and Sec.~\ref{sec2b}.
We then continue with the serial configuration defined by $\tilde \phi=0$, see Fig.~\ref{f1}(c) and Sec.~\ref{sec2c}. 
These two cases are expected to capture the essential Josephson physics in our setup.  
We finally allow for arbitrary phase configurations in Sec.~\ref{sec2d}, where we address the atomic limit for either 
the S lead ($\Delta_S\to\infty$) and/or the TS leads ($\Delta\to \infty$).

\subsection{Parallel configuration}\label{sec2b}

\begin{figure}[t]
\centering
\includegraphics[width=0.51\textwidth]{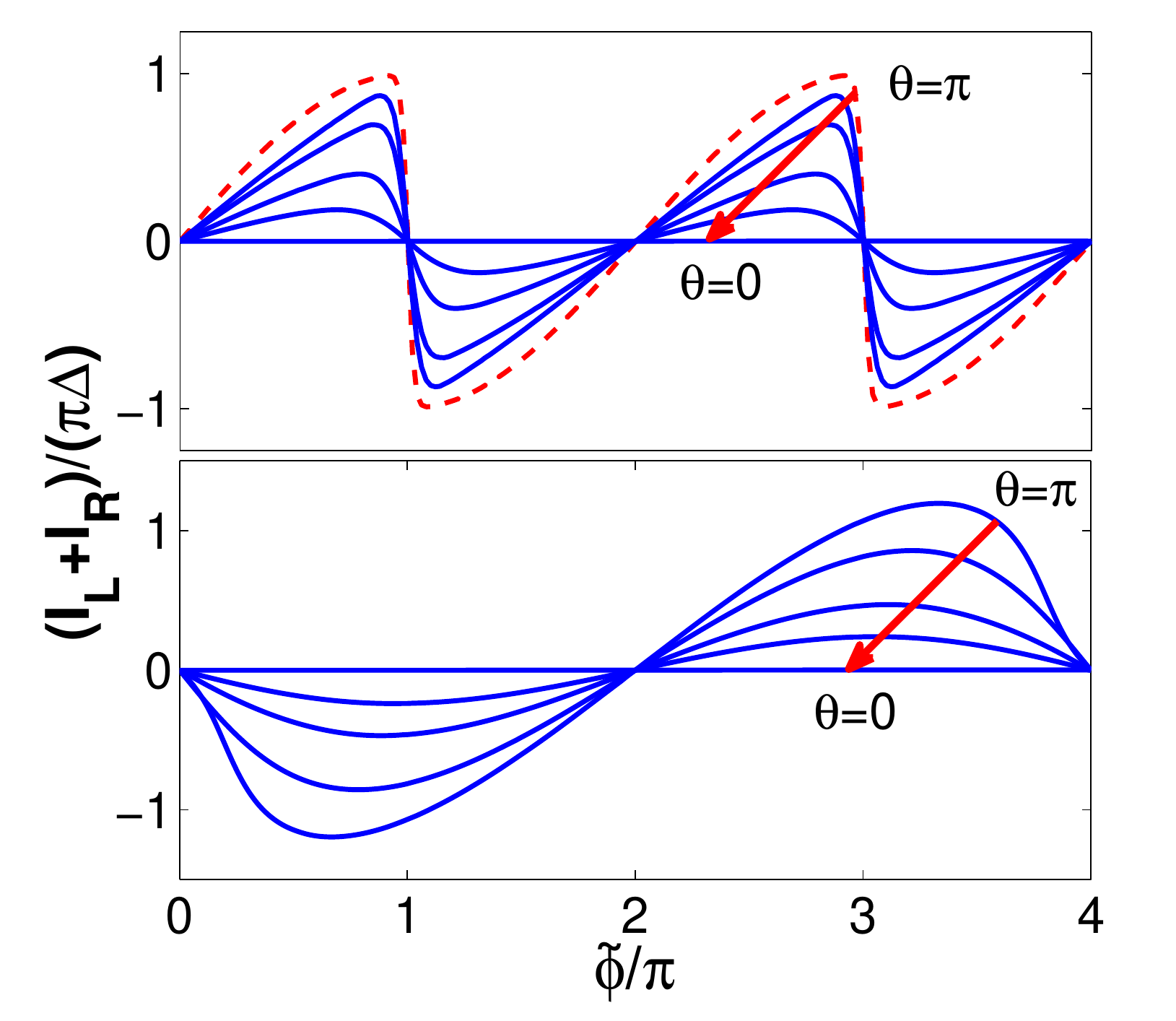}
\caption{\label{f2}
Supercurrent $I_L=I_R$ vs phase $\tilde \phi=\phi_L+\phi_R$ for the parallel configuration with
$\phi=\phi_L-\phi_R=0$ shown in Fig.~\ref{f1}(b), obtained numerically
from Eq.~\eqref{Ij} for $\lambda_L=\lambda_R=1$, $\Delta_S=\Delta$, and $T=0.02\Delta$.
From top to bottom, the spin angle is given by $4\theta/\pi=4, 3, 2, 1, 0$.  
The upper (lower) panel is for $\lambda_{LR}=0$ ($\lambda_{LR}=1$). The red dashed curve in the upper panel
is given by Eq.~(\ref{cpr2}) for $\Delta_S\to \infty$ and $\theta=\pi$. The blue curve for $\theta=\pi$ in the upper panel 
is also known analytically, see Eq.~\eqref{ana1}.}
\end{figure}

We begin with the parallel configuration shown in Fig.~\ref{f1}(b), where  $\phi=\phi_L-\phi_R=0$ and 
only $\tilde \phi=\phi_L+\phi_R$ can change.  In effect, we thus have a two-terminal Josephson junction with phase difference $\tilde \phi/2$
between a  two-channel TS lead and the S lead,  termed ``S-2TS junction'' in what follows.
Results for this configuration are shown in Fig.~\ref{f2} for various values of $\theta$.  
 
\subsubsection{Without direct TS-TS coupling}

Let us first discuss the case without direct TS-TS coupling, $\lambda_{LR}=0$, cf.~the upper panel in Fig.~\ref{f2}.
The critical current, $I_c={\rm max}\left|I_{L/R}(\tilde\phi)\right|$, has a maximum for antiparallel spin alignment ($\theta=\pi$)
and then monotonically decreases down to $I_c=0$ as $\theta$ is reduced to $\theta=0$.   
This behavior indicates that the supercurrent is  indeed due to Cooper pair splitting processes as anticipated in Sec.~\ref{sec1}.
Such processes are favored in the antiparallel spin configuration since a Cooper pair in the S lead is built from electrons with opposite spin.
On the other hand, they are ruled out for $\theta=0$. 
Remarkably, the CPR is $\pi$-periodic in the phase difference $\tilde \phi/2$ across the S-2TS junction. 

As discussed in Sec.~\ref{sec2d} below, analytical results for the CPR follow in the atomic limit for the S lead, $\Delta_S \to \infty$, where we find
\begin{eqnarray}\label{cpr2}
I_{L/R}^{(\Delta_S\to \infty)}(\tilde\phi)&=&  -\frac{\partial E_{A}}{\partial\tilde \phi}  \tanh(E_A/2T) ,\\ \nonumber
E_A (\tilde\phi)& = & \sqrt{\tau(\theta)} \Delta \cos(\tilde \phi/2).
\end{eqnarray}
The Andreev bound state (ABS) energy $E_A(\tilde\phi)$ depends on the spin canting angle $\theta$ via the transmission amplitude,
\begin{equation}\label{taudef}
\tau(\theta) = \frac{4 \lambda^4\sin^2(\theta/2) } {[1 + \lambda^4\sin^2(\theta/2)]^2},
\end{equation}
characterizing the transparency of the S-2TS Josephson junction with $\lambda_L=\lambda_R=\lambda$. 
The dependence of the ABS energy on the individual phases $\phi_{L/R}$ is $4\pi$-periodic, 
reflecting the fact that single electrons are transferred into/from each TS lead.  Nonetheless,
the actual CPR turns out to be $\pi$-periodic in the phase difference $\tilde\phi/2$ across the S-2TS junction, with jump-like
behavior of the CPR near $\tilde\phi = \pi~({\rm mod}~2\pi)$, cf.~the upper panel of Fig.~\ref{f2}. 
This peculiar behavior can be traced back to parity crossings and will be discussed in more detail in Sec.~\ref{sec2d}.

Another analytically accessible case is given by $\Delta_S=\Delta$. 
For instance taking $\theta = \pi$, where $I_c$ will be maximal, we obtain
\begin{equation}\label{ana1}
I_{L/R}^{(\Delta_S=\Delta)}(\tilde \phi) = -\sum_{\pm} 
\frac{ E^{(\pm)}_A }{\partial \tilde\phi} \tanh\left(\frac{E^{(\pm)}_A}{2T}\right),
\end{equation}
with a pair of ABS energies
\begin{equation}
E_A^{(\pm)}(\tilde \phi) = \frac{\Delta}{\sqrt{2}} \left( 1 \pm  \sqrt{1 - \tau^2 \cos^2 (\tilde \phi/2)} \right)^{1/2} ,
\end{equation}
where $\tau=4\lambda^2/(1+\lambda^2)^2$.   Equation \eqref{ana1} nicely matches the corresponding numerical result for $\Delta_S=\Delta$
shown in the upper panel of Fig.~\ref{f2}.  

\begin{figure}[t]
\centering
\includegraphics[width=0.53\textwidth]{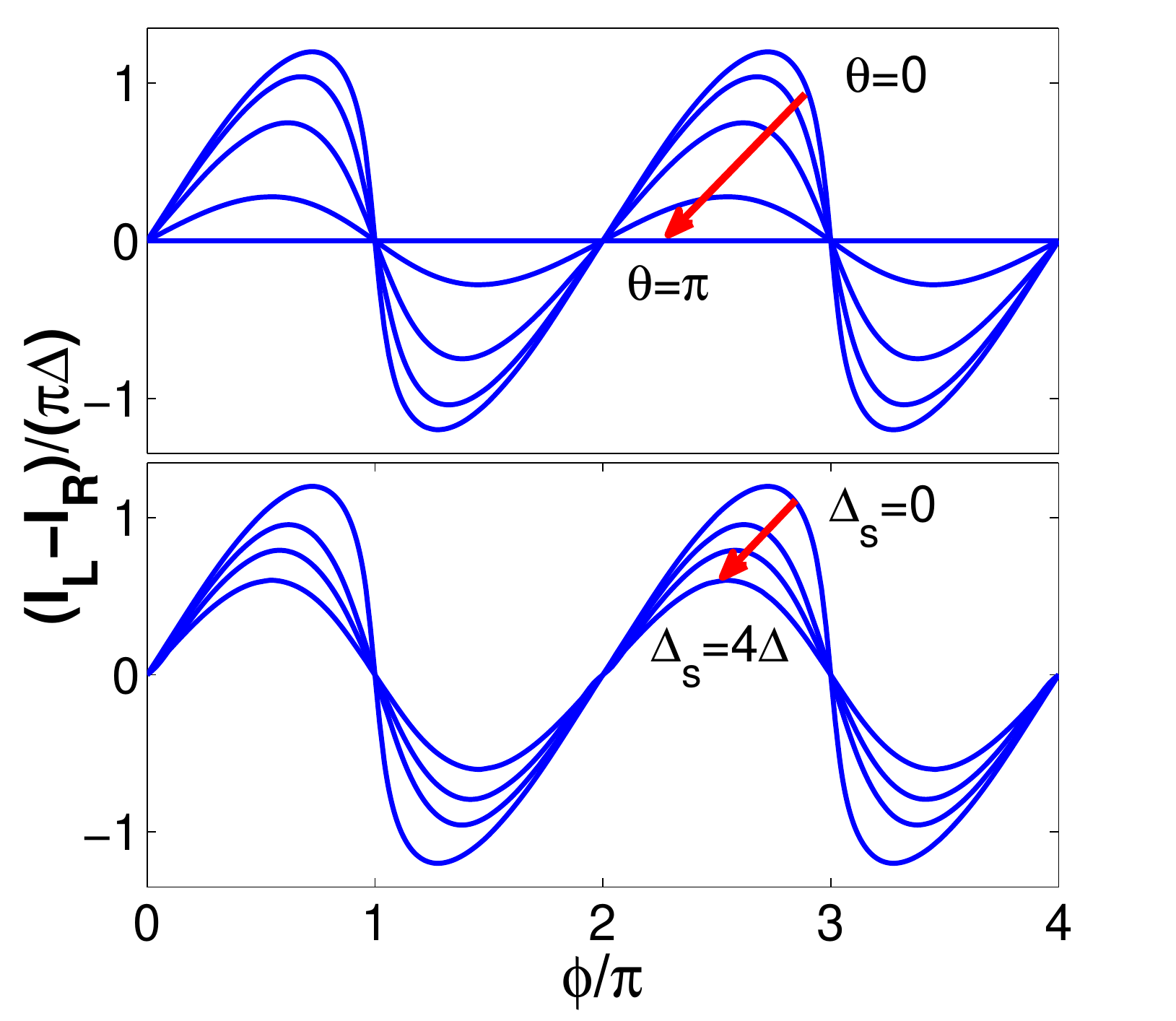}
\caption{\label{f3}
TS-TS supercurrent vs phase difference $\phi=\phi_L-\phi_R$ in the serial case $\tilde \phi=0$, see Fig.~\ref{f1}(c),
where $I_S=0$ and $I_L=-I_R$. The S lead acts as currentless probe.
The upper panel illustrates the supercurrent suppression with increasing spin angle $\theta$ for a normal probe ($\Delta_S = 0$), see Eq.~\eqref{tsnt},
for $4\theta/\pi=0, 1, 2, 3, 4$. 
The lower panel shows how current is suppressed with increasing $\Delta_S$ for $\theta = 0$, taking $\Delta_S/\Delta=0,1,2,4$.
All other parameters are as in Fig.~\ref{f2} with $\lambda_{LR} = 0$.}
\end{figure}

\subsubsection{With direct TS-TS coupling}

We next briefly address the effects of a direct tunnel coupling $\lambda_{LR}$ between both TS wires. Typical examples for the CPR 
are shown in the lower panel of Fig.~\ref{f2}, 
where, for simplicity, we assumed the $\theta$-independent value $\lambda_{LR}=1$.
This implies equal weights for TS-TS tunneling processes with and without spin flips.
Such an assumption can be phenomenologically 
justified by taking into account the magnetic field, cf.~the more microscopic 
model discussed in Sec.~\ref{sec3}.   

We observe from Fig.~\ref{f2} that $\lambda_{LR}$ has two important
consequences as compared to the case $\lambda_{LR}=0$ studied so far. 
First, the CPR becomes $2\pi$- rather than $\pi$-periodic.
Second, the CPR exhibits $\pi$-junction behavior, i.e., it appears with
 a phase shift of $\pi$ such that $\tilde\phi=0$ now represents a maximum of the junction free energy.  
Apart from the above two changes, however, the overall behavior of the CPR is similar to the case $\lambda_{LR}=0$.
In particular, the critical current is maximal for $\theta=\pi$ and vanishes for $\theta=0$.

The $\pi$-junction behavior observed numerically in Fig.~\ref{f2} can also be captured analytically in the  atomic limit $\Delta\to \infty$ for the TS wires, cf.~Sec.~\ref{sec2d}.
Indeed, for spin canting angle $\theta=\pi$ but arbitrary $\lambda_{LR}$, 
we obtain 
\begin{eqnarray}\nonumber
&& I_{L/R}^{(\Delta\to \infty)}=\frac{T}{2}\sum_\omega \left(
\lambda^2\sin\tilde\phi-2\lambda_{LR} \sqrt{1+\frac{\omega^2}{\Delta_S^2}} \sin(\tilde\phi/2)  \right)
 \\ \label{eqnew}
&& \quad \times 
\frac{\lambda^2\Delta^2_S}{\lambda^4\omega^2+[\lambda_{LR}\sqrt{\omega^2+\Delta_S^2}-\lambda^2\Delta_S\cos(\tilde\phi/2) ]^2}
\end{eqnarray}
with $\lambda_L = \lambda_R=\lambda$.
Evidently $\pi$-junction behavior, $\partial_{\tilde\phi} I_{L/R}(\tilde\phi=0)<0$, 
is predicted by Eq.~\eqref{eqnew} for 
sufficiently large direct TS-TS couplings, $\lambda_{LR} > \lambda^2$.

\subsection{Serial configuration}\label{sec2c}

We now turn to the serial configuration in Fig.~\ref{f1}(c), where $\phi_L=-\phi_R$ and hence $\tilde\phi=0$. 
This case can be viewed as a TS-TS Josephson junction with phase difference $\phi=\phi_L-\phi_R$
in the presence of a currentless (since $I_S=0$) probe electrode corresponding to the S lead.
Figure \ref{f3} shows numerical results for the CPR, $I_L(\phi)=-I_R(\phi)$, in this configuration.

\subsubsection{Without direct TS-TS coupling}

As in Sec.~\ref{sec2b}, we start our analysis with the case $\lambda_{LR}=0$. A finite supercurrent 
flowing between the TS leads can then only be mediated through (real or virtual) tunneling processes via states in the S lead. 
Note that such states are only available at energies above $\Delta_S$.  This suggests that the critical current $I_c$ will
be maximal for $\Delta_S=0$ and then decreases with increasing $\Delta_S$.  Furthermore, since tunneling processes
between the TS leads have highest amplitude for parallel spin alignment, we expect that $I_c$ 
is now maximal for $\theta=0$ but then will be suppressed with increasing $\theta$.  We emphasize that the $\theta$-dependence of
$I_c$ is reversed with respect to the parallel configuration in Sec.~\ref{sec2b}.    The current
suppression with increasing $\theta$ and/or $\Delta_S$ is shown in the upper and lower panel of Fig.~\ref{f3}, respectively.
It is worth mentioning that the CPR, $I_L(\phi)$, is always $2 \pi$-periodic.

In order to better understand the results in Fig.~\ref{f3}, we again turn to analytically accessible limits. In particular,
when the S lead has a vanishing gap ($\Delta_S=0$) and thus represents a normal-conducting probe, Eq.~\eqref{Ij} 
can be simplified to the expression
\begin{eqnarray}  \label{tsnt}
I_L^{(\Delta_S=0)} (\phi) &= &- T \sum_\omega \frac{2 E_A \partial_\phi E_A}{\omega^2 + E_A^2 + \gamma |\omega| \sqrt{\omega^2 + \Delta^2}},\\
\nonumber
E_A(\phi) &=& \tilde \Delta \sqrt{1 - \tilde\tau \sin^2 (\phi/2)} ,
\end{eqnarray}
with 
\begin{eqnarray}
\tilde \Delta &=& \frac{\Delta}{\sqrt{1+y^2}},\quad \tilde\tau=\cos^2(\theta/2), \\
\nonumber 
y&=& \frac{1 + \lambda^4 \sin^2 (\theta/2)}{ 2 \lambda^2}.
\end{eqnarray}
Importantly, the rate  $\gamma = 2 y/(1 + y^2)$ in Eq.~\eqref{tsnt} 
 causes ABS damping even though the $2\pi$-periodic ABS energy $E_A$ is always detached from the TS continuum,
$E_A<\Delta$.  Typical CPR curves resulting from Eq.~\eqref{tsnt} are shown in the upper panel of Fig.~\ref{f3}.

\begin{figure}[t]
\centering
\includegraphics[width=0.52\textwidth]{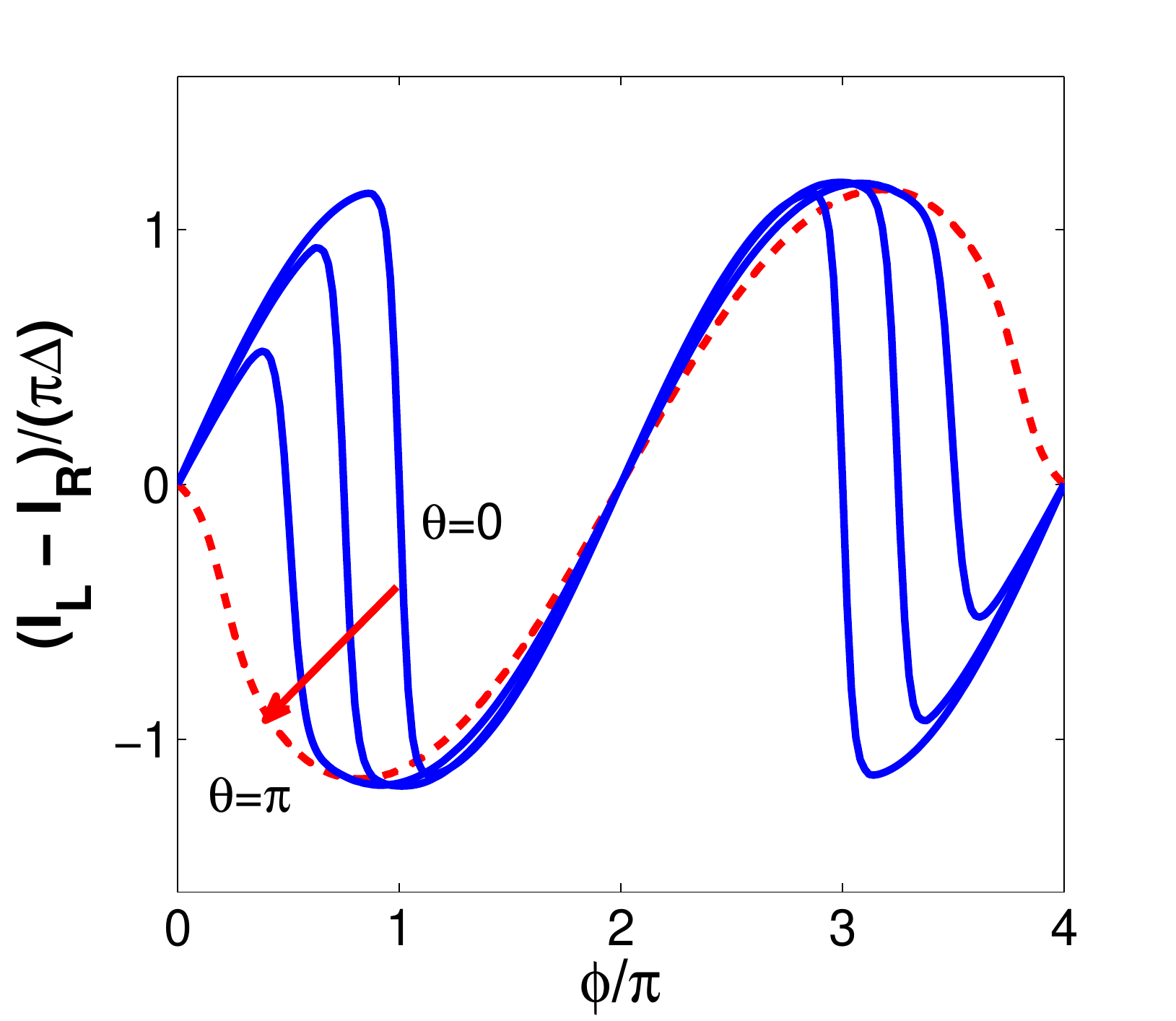}
\caption{\label{f4}
TS-TS supercurrent vs $\phi$ in the serial configuration ($\tilde \phi=0$) obtained from Eq.~\eqref{Ij} for direct TS-TS coupling $\lambda_{LR}=1$
and several values of the spin canting angle, $4\theta/\pi=0,1,2,4.$
All other parameters are as in Fig.~\ref{f3}.}
\end{figure}

Analytical progress is also possible for $\Delta_S = \Delta$, where  no damping terms are encountered.
Focusing on the case of maximal supercurrent, $\theta = 0$, we then obtain
\begin{equation}\label{delsdel}
I^{(\Delta_S=\Delta)}_L(\phi) =-\frac{\partial E_A}{\partial \phi} \tanh\left(\frac{E_A }{ 2T}\right) ,
\end{equation}
with the ABS energy
$E_A(\phi) = \Delta \sqrt{1 - \tau_1 \sin^2 (\phi/2)}$ and $\tau_1 = 4 \lambda^4/(1 + 2 \lambda^2)^2.$
Equation \eqref{delsdel} matches the numerical curve with $\Delta_S=\Delta$ and $\theta=0$ shown in the 
lower panel of Fig.~\ref{f3}, and therefore illustrates the
supercurrent suppression with increasing $\Delta_S/\Delta$ at fixed spin angle $\theta$.

\subsubsection{With direct TS-TS coupling}

In the presence of a direct TS-TS coupling $\lambda_{LR}$, the $2\pi$-periodicity of the CPR found above for $\lambda_{LR}=0$ 
may turn into a $4\pi$-periodicity.  This remarkable effect is shown in Fig.~\ref{f4}.
Such a periodicity is found in the absence of coupling to the S lead ($\lambda_{L/R}=0$) when the parity of the junction is conserved \cite{Alicea2012}.
In our setup, the $4\pi$-periodicity  can be understood by noting that the S lead acts as a probe electrode and thereby effectively fixes
the fermion parity of the trijunction.
As illustrated in Fig.~\ref{f4}, we find that the $4\pi$ Josephson effect takes place for all $\theta\ne 0$ (mod~$2\pi$).  

The $4\pi$-periodicity can also be confirmed analytically in the atomic limit
$\Delta\to \infty$ for the TS wires. For $\theta=\pi$ and 
arbitrary $\lambda_{LR}$, we then obtain 
\begin{eqnarray}\nonumber
&& I_{L}^{(\Delta\to \infty)}=\frac{T}{2}\sum_\omega \left(
\lambda_{LR}\sin\phi-\frac{2\lambda^2\Delta_S}{\sqrt{\omega^2+\Delta_S^2}} 
\sin(\phi/2)  \right)
 \\ \label{eqnw1}
&& \quad \times 
\frac{\lambda_{LR}(\omega^2+\Delta^2_S)}
{ \lambda^4\omega^2+[\lambda_{LR}\sqrt{\omega^2+\Delta_S^2} 
\cos(\phi/2)-\lambda^2\Delta_S ]^2 }
\end{eqnarray}
with $\lambda_L = \lambda_R=\lambda$.  This result explicitly shows the
 $4\pi$-periodicity of the CPR.  Note that in addition, $\pi$-junction behavior is 
 predicted by Eq.~\eqref{eqnw1} in the parameter regime $\lambda_{LR}<\lambda^2$. 
 Let us also mention in passing that 
when the atomic limit is also taken for the S lead, $\Delta_S\to \infty$, 
Eq.~\eqref{eqnw1} simplifies to Eq.~\eqref{cprserial} below (taken for $\theta=\pi$).

 It is worth stressing that the TS-S-TS setup thereby provides a completely robust strategy to measure the $4\pi$-periodic CPR 
 expected in the ground state of a TS-TS junction with conserved fermion parity \cite{Kitaev2001,Alicea2012,Leijnse2012,Beenakker2013}.  
This setup could therefore significantly simplify future experimental  studies of the topological Josephson effect in a true equilibrium state.

\subsection{Atomic limit}\label{sec2d}

We now address  arbitrary phase configurations $\phi_{L/R}$  but restrict ourselves to cases where the TS gap $\Delta$ and/or
the S gap $\Delta_S$ become much larger than all other energy scales of interest.  Effectively taking $\Delta\to \infty$ and/or 
$\Delta_S\to \infty$ defines the atomic limit for the TS/S leads, where one can approximate the respective lead
by a zero-bandwidth model.

First, for $\Delta_S\to \infty$, the self-energy $\Sigma(\omega)$ in Eq.~\eqref{selfen}  
simplifies to $\Sigma_{LL/RR}=0$ while $\Sigma_{LR} (\omega)=\Sigma_{RL}^\dagger(-\omega)$
 becomes frequency independent and hence local in time,  
\begin{equation}\label{selfen-atomic}
\Sigma_{LR}=\left( \begin{array}{cc}\lambda_{LR}  e^{-i \phi/2} & -\lambda_L\lambda_R\sin(\theta/2) e^{-i \tilde \phi/2} \\
\lambda_L\lambda_R\sin(\theta/2)  e^{i \tilde \phi/2} & -\lambda_{LR} e^{i \phi/2} \end{array} \right),
\end{equation}
where we  use Eq.~\eqref{phasedef}.

For $\Delta\to \infty$, on the other hand, above-gap excitations in the TS wires become negligible and a projection to the
respective low-energy MBS sector 
is sufficient. Taking this limit in Eq.~\eqref{bGFK} with $\Delta_L=-\Delta_R=\Delta$, we obtain the TS bGFs,
$G_{L/R}(\omega) = -i(\Delta/\omega) [\sigma_0\pm \sigma_x].$
Introducing Majorana operators $\gamma_{L,R}^{}= \gamma_{L,R}^\dagger$ with anticommutator $\{ \gamma_j,\gamma_{j'}\}=\delta_{jj'}$, 
this limiting form of the bGF implies a
projection of the boundary spinors $\Phi_j$ in Sec.~\ref{sec2a} onto the subspace spanned by MBSs only,
\begin{equation}\label{phiproj}
\Phi_L\to \sqrt{\Delta}\left(\begin{array}{c} 1\\ 1\end{array}\right) \gamma_L, 
\quad
\Phi_R\to i\sqrt{\Delta}\left(\begin{array}{c} 1\\ -1\end{array}\right) \gamma_R.
\end{equation}
Taking also the atomic limit for the S lead, the time-local form of $\Sigma(\omega)$ in Eq.~\eqref{selfen-atomic}
implies that the Dyson equation \eqref{dyson} is solved by diagonalizing the effective Hamiltonian 
\begin{eqnarray}\label{atomicli}
H_{at} &=&  \frac12 \Phi_L^\dagger \Sigma_{LR} \Phi_R+ {\rm h.c.}=iE_A(\phi,\tilde\phi)\gamma_L\gamma_R,\\
E_A &=& 2\Delta\left( \lambda_{LR}\cos(\phi/2) + \lambda_L\lambda_R\sin(\theta/2) \cos(\tilde\phi/2)\right).\nonumber
\end{eqnarray}
Note that Eq.~\eqref{atomicli} recovers the ABS energy \eqref{cpr2} for $\lambda_{L,R}\ll 1$ and $\lambda_{LR}=0$.
Moreover, for $\lambda_{L/R}=0$,  Eq.~\eqref{atomicli} reduces to the celebrated $4\pi$-periodic ABS energy of a two-terminal TS-TS junction
  \cite{Kitaev2001,Alicea2012,Leijnse2012,Beenakker2013}.   

  Importantly,  Eq.~\eqref{atomicli} provides insights concerning
  $\pi$-junction behavior.  To that end, let us consider the serial configuration in 
 Sec.~\ref{sec2c}. Putting $\tilde\phi=0$ in Eq.~\eqref{atomicli}, we obtain the $T=0$ CPR 
 \begin{eqnarray}\label{cprserial}
 I_L(\phi)&=& -I_R=\frac{e\lambda_{LR}\Delta}{\hbar} \sin(\phi/2) \times \\ \nonumber
 & & \qquad \times \quad {\rm sgn}\left[\lambda_{LR}\cos(\phi/2)+\lambda_L\lambda_R\sin(\theta/2)\right].
 \end{eqnarray}
The CPR is $4\pi$-periodic and will show a transition to the $\pi$-junction regime once the sign factor ${\rm sgn}[\cdots]$ becomes negative.
 In fact, $\pi$-junction behavior is expected to be rather common in TS-S-TS trijunctions near the atomic limit.

\section{Boundary GF approach for spinful nanowires}\label{sec3}

So far our description of the TS leads has been based on the low-energy limit of a semi-infinite Kitaev chain, see Sec.~\ref{sec2}.
A more microscopic approach tailor-made for semiconductor nanowire implementations is discussed in this section.
To that end, we first derive the bGF for the spinful TS nanowire model of Refs.~\cite{Lutchyn2010,Oreg2010}, which 
arguably can give a rather accurate description of the band structure.
This model applies for a single conduction channel but includes spin-orbit coupling, proximity-induced pairing, and a magnetic Zeeman field.
In a second step, we then apply this bGF to the problem of Josephson transport
in TS-S-TS junctions and compare the results to those in Sec.~\ref{sec2}.

\subsection{Derivation of the bGF}\label{sec3a}

A spatially discretized version of the model in Refs.~\cite{Lutchyn2010,Oreg2010} is given by
\begin{eqnarray}\label{oregmodel}
H_{\rm wire} &=&\frac12 \sum_j \left[\psi_j^{\dagger} \hat{h} \psi_j^{} + 
\left( \psi_j^{\dagger} \hat{t} \psi_{j+1}^{}  + \mbox{h.c.}\right)\right],\\
\nonumber 
\hat{h} &=& (2t-\mu)\sigma_z \tau_0 + V_x \sigma_0\tau_x + 
\Delta \sigma_x\tau_0,\\ \nonumber
\hat{t} &=& -t\sigma_z\tau_0 + i\alpha\sigma_z\tau_z,
\end{eqnarray}
where the fermion operators $c_{j\sigma}$ for a given lattice site $j$ 
and spin $\sigma$ are combined in the
 four-spinor $\psi^T_j= \left(c_{j\uparrow}^{}, 
c_{j\downarrow}^{}, c^{\dagger}_{j\downarrow}, -c^{\dagger}_{j\uparrow} \right)$. 
In Eq.~\eqref{oregmodel}, the Pauli matrices $\sigma_{x,y,z}$ as well as the identity $\sigma_0$ act as before in Nambu (particle-hole) space, while 
Pauli matrices $\tau_{x,y,z}$ and the identity $\tau_0$ are defined in spin space.

To select reasonable model parameters for the calculations below, we first choose the lattice spacing $a=10$~nm,
which yields the nearest-neighbor hopping element $t = \hbar^2/(2m^*a^2)$ 
 and the spin-orbit coupling $\alpha=\hbar u/a$,  where $u$ is the spin-orbit interaction parameter in the continuum model 
 of Ref.~\cite{Oreg2010}.  With the effective mass $m^*\simeq 0.02 m_e$ for InAs, we obtain $t=20$~meV 
 and estimate $\alpha\simeq 4$~meV from the measurements in Ref.~\cite{Schapers2010}. 
 The Zeeman scale $V_x$ is determined by the magnetic field $B$ along the wire,
$V_x = \mu_B g B/2$, with Land{\'e} factor  $g\approx 10$ in InAs. 
For the gap parameter, we assume $\Delta=0.2$~meV.  The remaining free variables are $B$ (and hence $V_x$) and
the chemical potential $\mu$, which are  key experimental parameters for changing the nanowire properties.

\begin{figure}[t]
\begin{center}
\includegraphics[width=0.49\textwidth]{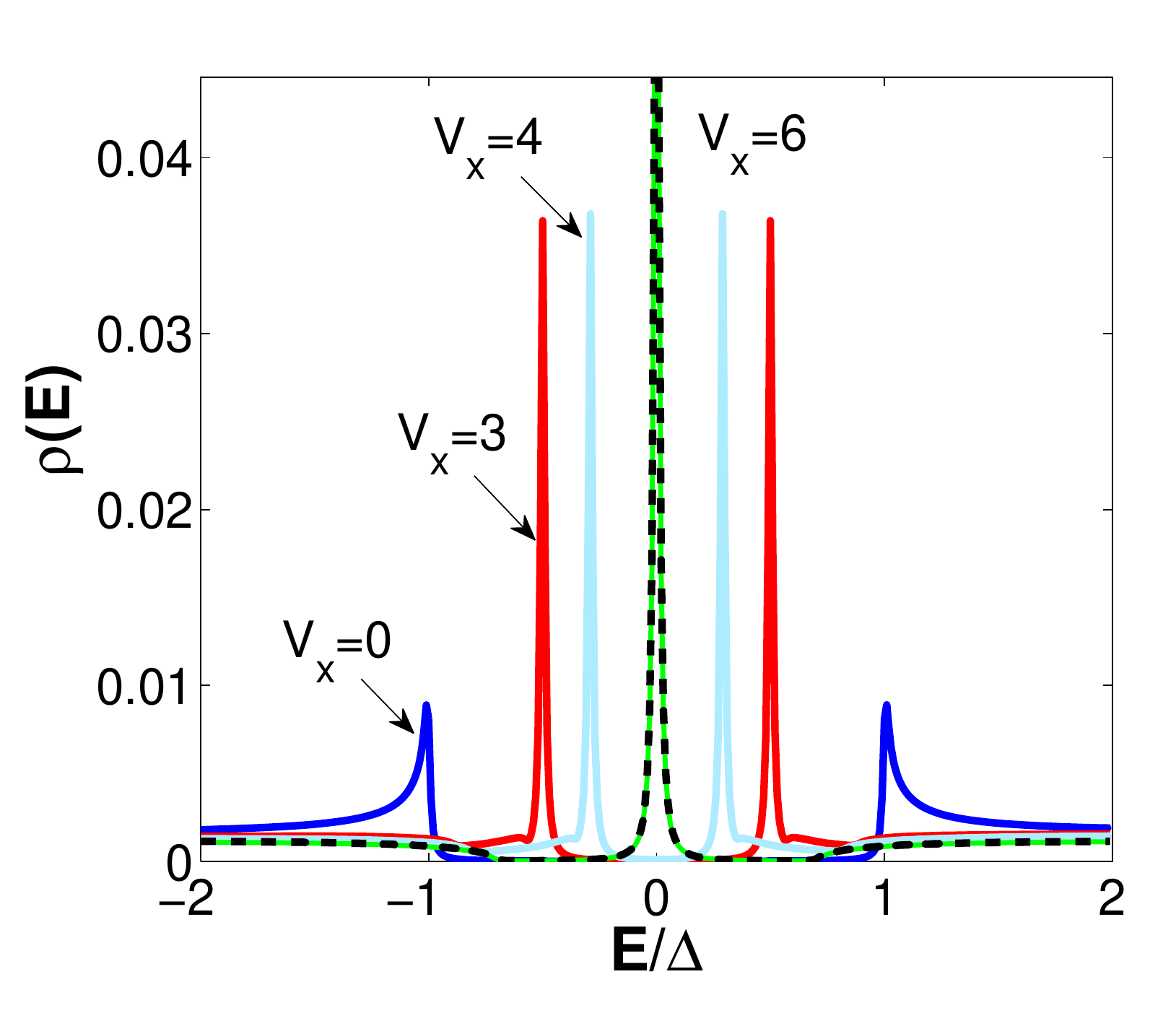}
\end{center} 
\caption{Local density of states $\rho(E)$ (in units of $1/\Delta$) at the wire boundary vs energy $E$ (in units of $\Delta)$, 
using the spinful wire model \eqref{oregmodel}.  We have used a broadening parameter $\gamma_0=0.01\Delta$  
to visualize the subgap structure.
We put $\mu=5$~meV and vary the Zeeman scale $V_x$ from 0 to $6$~meV, 
with all other parameters specified in the main text.  The topological
phase is realized for $V_x>V_c=\sqrt{\mu^2+\Delta^2}\simeq 5.004$~meV \cite{Lutchyn2010,Oreg2010}. 
For $V_x=6$~meV, the zero-energy Majorana peak is observed. The dashed curve gives the corresponding result
obtained from the Kitaev model in Sec.~\ref{sec2}, with an effective proximity gap and an effective normal density of states.
} 
\label{f5}
\end{figure}

In this section, we work with the retarded GF $G^{r}(\omega)$ 
of a given TS wire, which is obtained 
by analytic continuation, $i\omega\to \omega+i 0^+,$ 
from the respective Matsubara GF $G(\omega)$
discussed in Sec.~\ref{sec2}.  To lighten notation, we drop the $r$ 
superscript from now on. 
For an infinitely long nanowire described by Eq.~\eqref{oregmodel},
the GF in spin-Nambu space (indicated by the hat superscript)  follows in 
frequency-momentum representation as
\begin{eqnarray}\label{gff0}
\hat{G}^{(0)}(k,\omega) &=& \left[\omega - \hat{h} - \hat{V}(k)\right]^{-1},\\
\hat{V}(k) &=& -2t\cos(ka)\sigma_z\tau_0 +2 \alpha\sin(ka)\sigma_z\tau_z.\nonumber
\end{eqnarray}
Explicit expressions for the 
matrix elements  $[\hat G^{(0)}(k,\omega)]_{lm}$ 
in spin-Nambu space are summarized in Appendix \ref{appa}.
The corresponding real-space representation is obtained by contour integration,
\begin{equation}
\hat{G}^{(0)}_{lm}(\omega) = \oint_{|z|=1} \frac{dz}{iaz} \hat{G}^{(0)}(k,\omega)
z^{(l-m)},\quad z=e^{ika}.
\end{equation}
 We then determine the poles of $\hat{G}^{(0)}(k,\omega)$ in the $z$-plane numerically, 
while the corresponding residues are computed analytically. 
Finally, the bGF for the left/right semi-infinite spinful TS wire follows by solving the
respective Dyson equation for breaking the infinite chain \cite{Zazunov2016}, 
\begin{eqnarray}
\hat{G}_L &=& \hat{G}^{(0)}_{00} - \hat{G}^{(0)}_{01} \left(\hat{G}^{(0)}_{00}\right)^{-1} \hat{G}^{(0)}_{10}, \nonumber\\
\hat{G}_R &=& \hat{G}^{(0)}_{00} - \hat{G}^{(0)}_{10} \left(\hat{G}^{(0)}_{00}\right)^{-1} \hat{G}^{(0)}_{01},\label{gfb}
\end{eqnarray}
where the indices refer to lattice sites.

As a first example for this approach, we study the local density of states at the boundary of the wire,
$\rho(E)=-(1/\pi) {\rm Im}  [\hat G_{L}(E)]_{11}$,  where the indices now
refer to spin-Nambu space.
Figure~\ref{f5} illustrates the evolution of $\rho(E)$ as one varies the Zeeman scale $V_x$. 
Below the critical value $V_c=\sqrt{\mu^2+\Delta^2}$,  the topologically trivial phase without MBSs is realized. 
For $V_x>V_c$, however, Majorana zero modes develop and cause a zero-energy peak in $\rho(E)$. 
 For comparison, Fig.~\ref{f5} also  shows the limiting behavior predicted by the Kitaev model used in Sec.~\ref{sec2}.

\subsection{Spin structure of MBSs}

\begin{figure}[t]
\centering
\includegraphics[width=0.48\textwidth]{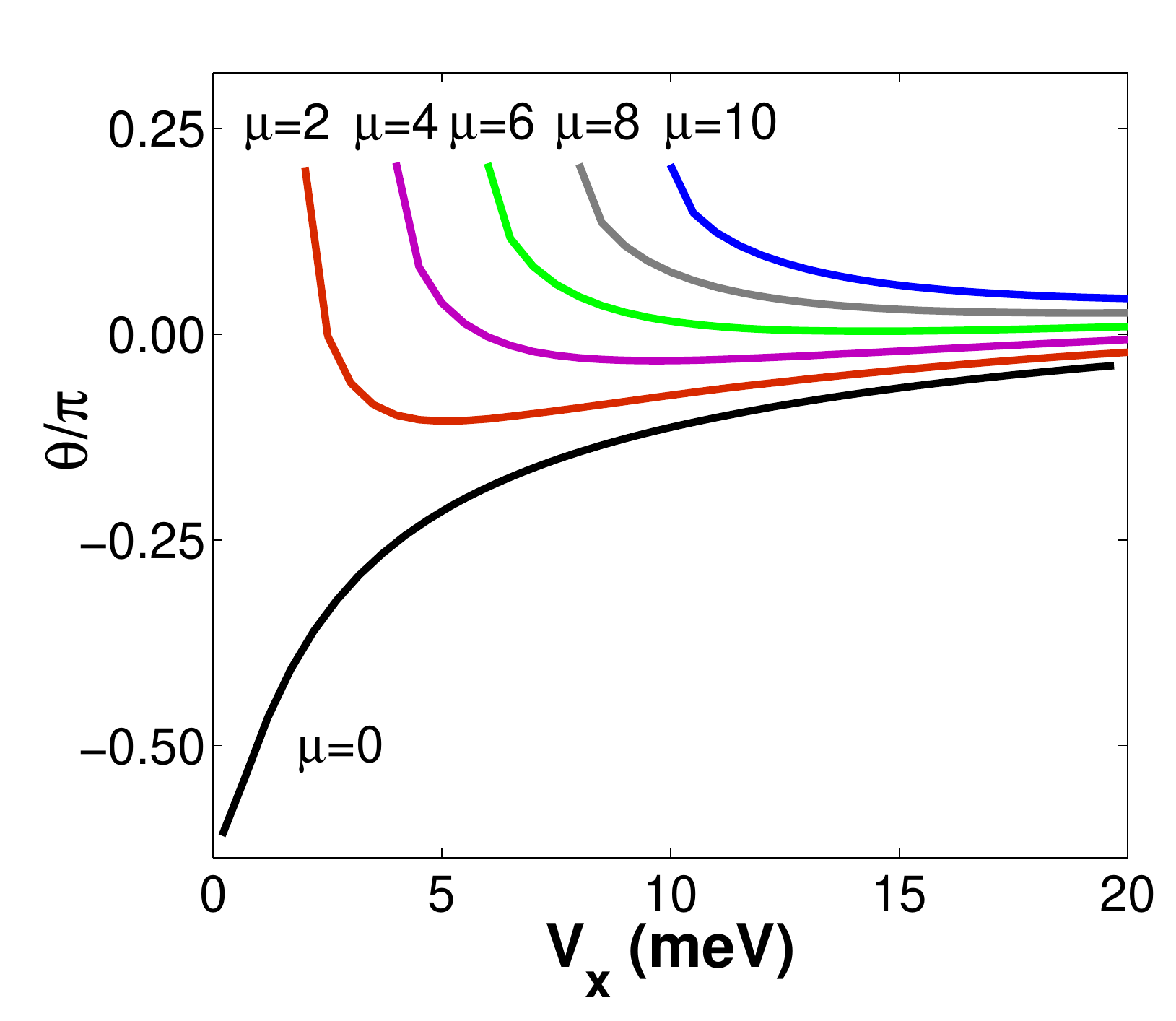}
\caption{Spin canting angle $\theta=\theta_L-\theta_R$ vs Zeeman scale $V_x$ obtained from Eq.~\eqref{thj} for different values of $\mu$.
The shown results are within the topological regime, $V_x>V_c$. All other parameters of the spinful TS wire model \eqref{oregmodel} are specified in the main text. } 
\label{f6}
\end{figure}

The spinful model \eqref{oregmodel} allows us to analyze the MBS spin structure and its evolution under changes of the
 magnetic field and/or the chemical potential. 
The spin spectral density at the boundary of the 
respective TS wire has the components ($\nu=x,y,z$)
\begin{equation}\label{spinden}
S_{\nu,L/R}(\omega) = \frac{1}{2\pi i} \mbox{Tr}
\left[\left(1+\sigma_z\right)\tau_\nu\left(G^a_{L/R}(\omega) 
-G^r_{L/R}(\omega)\right)\right] ,
\end{equation}
with  a trace over spin-Nambu space, and the advanced GF $G^a$.
As discussed in Refs.~\cite{Flensberg2011,Sticlet2012,Prada2012}, 
the MBS spin polarization points within a plane
perpendicular to the spin-orbit direction (the $x$-$y$ plane in our case). We may 
then determine the spin angles
in Fig.~\ref{f1}(a) from the zero-energy limit of Eq.~\eqref{spinden},
\begin{equation}\label{thj}
\tan \theta_{L/R} = \lim_{\omega\to 0} \frac{S_{y,L/R}(\omega)}{S_{x,L/R}(\omega)}.
\end{equation}
The symmetry of the setup implies $\theta_R=-\theta_L$. Hence we can compute the spin canting angle $\theta=\theta_L-\theta_R$,
which was introduced in Sec.~\ref{sec2} on purely phenomenological grounds, as a function of the model parameters in Eq.~\eqref{oregmodel}.  

The resulting dependence of $\theta$ on $V_x$ and/or $\mu$ is shown in Fig.~\ref{f6}. 
With increasing Zeeman scale $V_x$ (always in the topological regime $V_x>V_c$), we find that $\theta$
decreases as a power law where the exponent  depends on system parameters, see also Ref.~\cite{Prada2012}. 
This decrease becomes less pronounced for large values of $\mu$ which may therefore be advantageous
in achieving a widely tunable spin angle.
In any case, the choice of $V_x$ and/or $\mu$ will affect $\theta$ as shown in Fig.~\ref{f6}, which in turn affects
 Josephson transport in the trijunction as described in Sec.~\ref{sec2}.

\subsection{CPR across the topological transition}

\begin{figure}[t]
\centering
\includegraphics[width=0.52\textwidth]{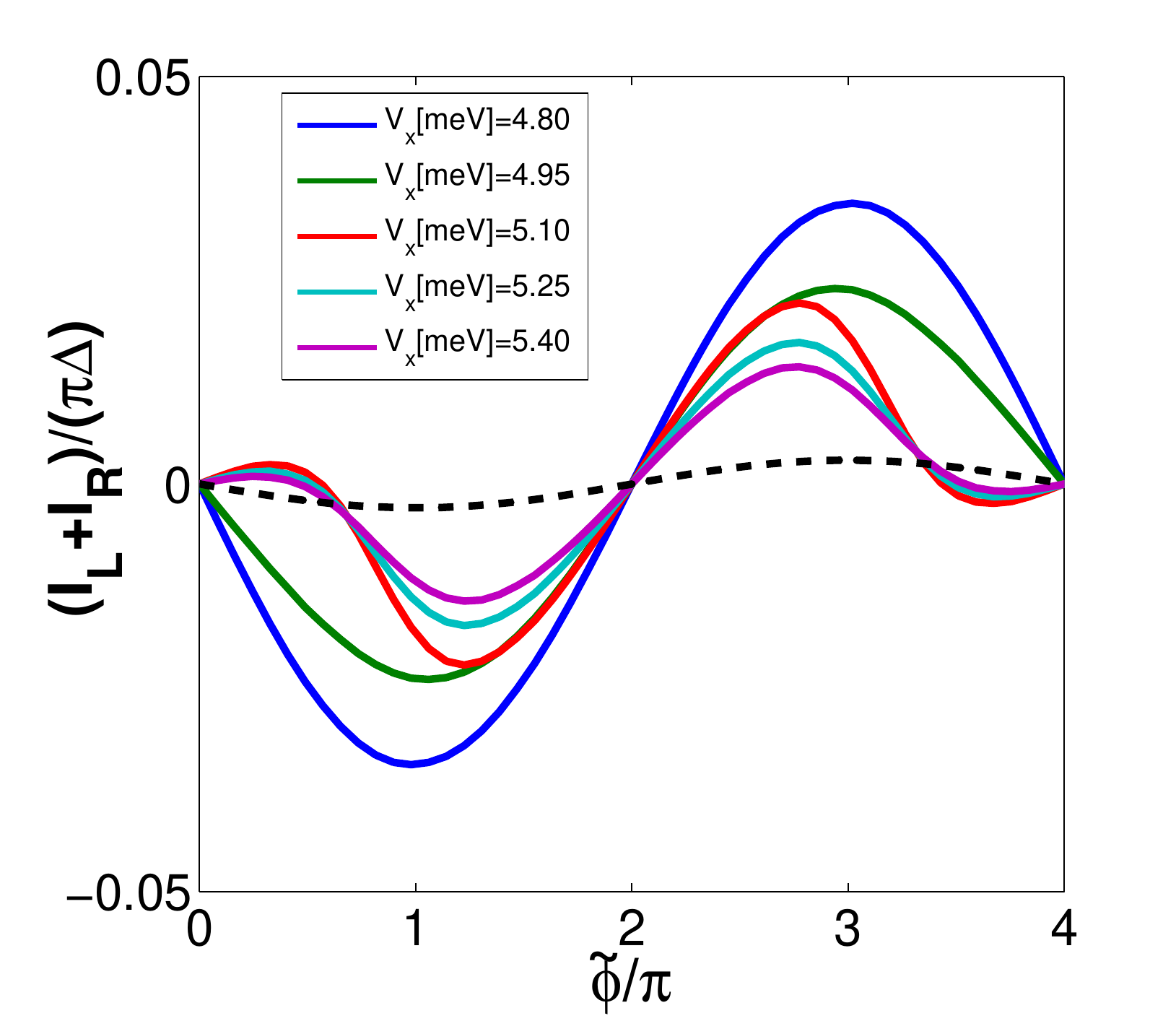}
\caption{CPR for the parallel configuration from the spinful model without direct TS-TS coupling, $\lambda_{LR}=0$.
The Zeeman scale $V_x$ is ramped through the topological transition at $V_c\simeq 5.004$~meV, with $\mu=\lambda_L=\lambda_R=5$~meV. 
The dashed curve gives the CPR for $\lambda_R=0$ and $V_x=5.4$~meV.  For all other parameters, see main text.} 
\label{f7}
\end{figure}

We next use the spinful model for analyzing the trijunction CPR across the topological transition at $V_x=V_c$. 
The tunnel Hamiltonian (\ref{Ht}) can be directly extended to the spinful case. However,
we do not absorb density of states factors in the tunnel amplitudes $\lambda_{L,R,LR}$ anymore (as in Sec.~\ref{sec2})
but instead specify them in meV units.

Let us first study the parallel configuration $\phi=0$, see Fig.~\ref{f1}(b), for the case
without direct TS-TS coupling, $\lambda_{LR}=0$.  We also put $\Delta=\Delta_S$ even though the effective TS gap becomes
reduced when crossing the topological transition, cf.~Fig.~\ref{f5}.  
The resulting CPR is shown in Fig.~\ref{f7} for $\mu=5$~meV and $\lambda_L=\lambda_R=5$~meV. 
We observe a transition from the conventional $2\pi$-periodic behavior for $V_x<V_c$ to a distorted CPR for $V_x>V_c$ that 
gradually resembles the CPR predicted by the Kitaev model in Sec.~\ref{sec2}, cf.~the upper panel in Fig.~\ref{f2}.
We emphasize that in order to match to results obtained from the Kitaev model, one 
should have $\mu \gg \Delta$. For $\mu \to 0$, the $s$-wave pairing contribution turns out to be important and masks the
corresponding behavior. For reference, the dashed curve in Fig.~\ref{f7} specifies the S-TS Josephson current for 
$\lambda_R=\lambda_{LR}=0$, where the right TS wire is completely decoupled. 
The  CPR should then vanish according to the model in Sec.~\ref{sec2}, see Ref.~\cite{Zazunov2012}.
However, when using the spinful model, we observe a finite (albeit much reduced) critical current.

\begin{figure}[t]
\centering
\includegraphics[width=0.52\textwidth]{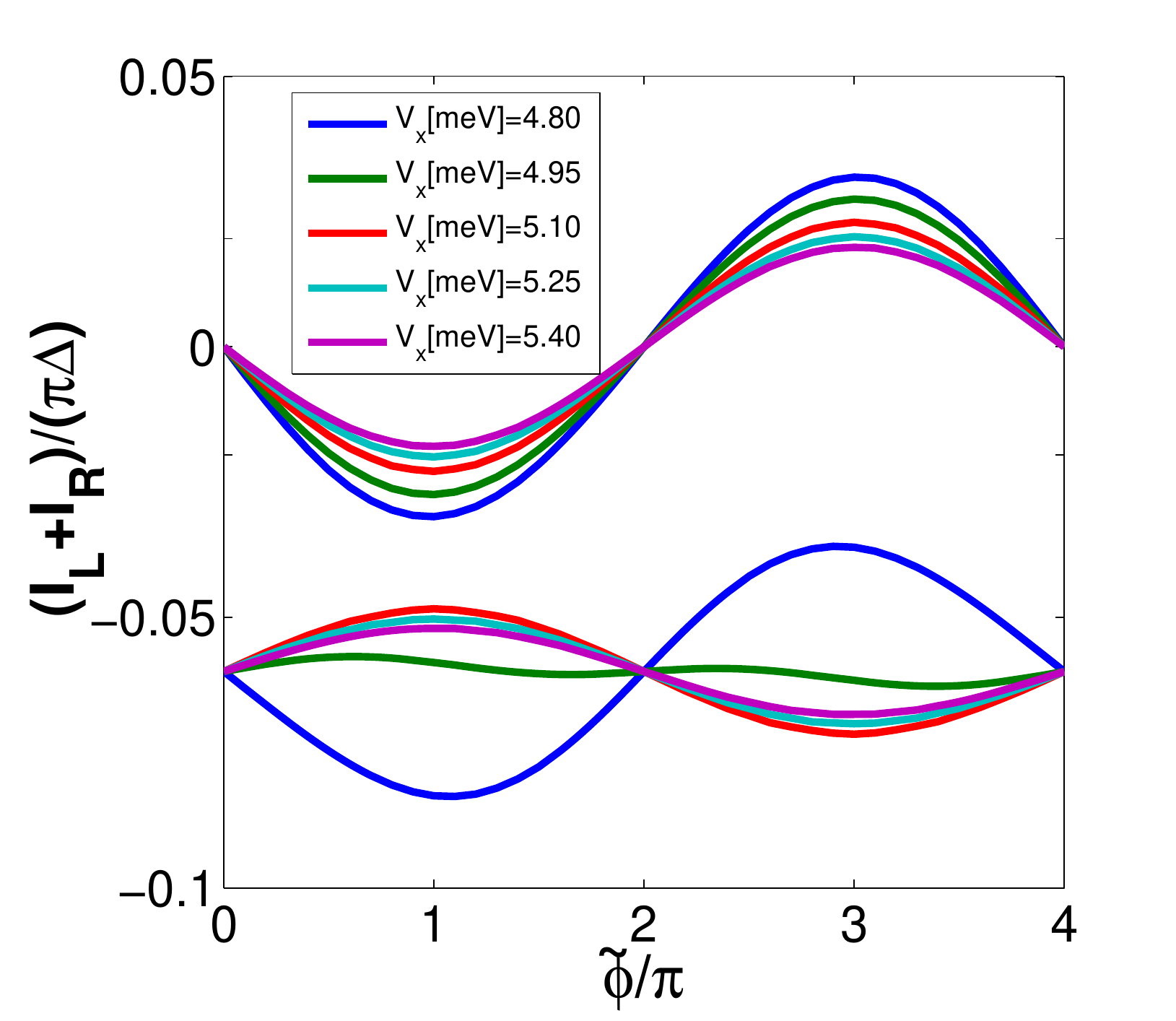}
\caption{Same as in Fig.~\ref{f7} but including a direct TS-TS coupling, $\lambda_{LR}\ne 0$.
The lower (upper) set of curves corresponds to $\lambda_{LR}=5$~meV ($\lambda_{LR}=-5$~meV),
where the lower curves have been shifted by $-0.06$ for clarity.} 
\label{f8}
\end{figure} 

Next, Fig.~\ref{f8} shows results for the same configuration as in Fig.~\ref{f7} but allowing for a direct TS-TS coupling $\lambda_{LR}$. 
We observe that the behavior of the CPR across the transition critically depends on the sign of $\lambda_{LR}$. 
When $\lambda_{LR} <0$, the CPR exhibits  $\pi$-junction 
behavior both for $V_x<V_c$ and $V_x>V_c$, and thus the evolution across the transition is smooth. However, for $\lambda_{LR} >0$, 
a change from $\pi$-junction to $0$-junction behavior occurs right at the topological transition.  This sensitivity to the sign of the direct tunnel 
matrix element $\lambda_{LR}$  is characteristic for our topological TS-S-TS trijunction and would not be found in topologically trivial systems. 
  
\begin{figure}[t]
\centering
\includegraphics[width=0.5\textwidth]{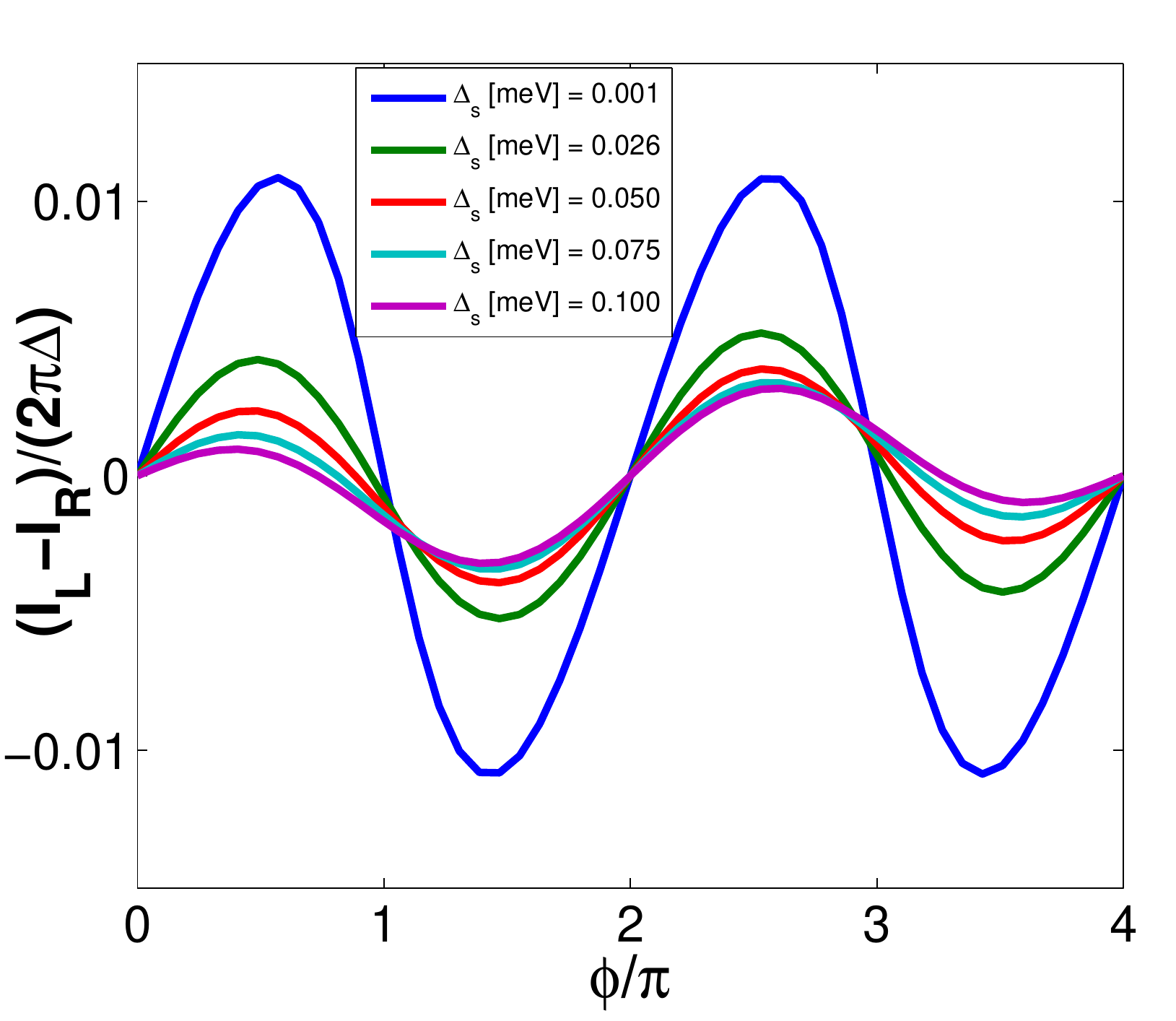}
\caption{CPR for the serial configuration from the spinful model for several values of $\Delta_S$. Parameters were chosen as
 $\mu=\lambda_L=\lambda_R=5$ meV, $V_x=7$ meV, and $\lambda_{LR}=0$. For all other parameters, see main text.} 
\label{f9}
\end{figure} 

Finally, we have also explored the CPR for the spinful model in the serial configuration, cf.~Fig.~\ref{f1}(c). Our results in Fig.~\ref{f9}
correspond to a situation where both TS leads are deep in the topological phase. 
As a consequence, the spin angle is quite small, 
cf.~Fig.~\ref{f6}, and hence large supercurrents
are expected from Sec.~\ref{sec2c}, cf.~the upper
panel of Fig.~\ref{f3}.
Next, Figure \ref{f9} illustrates the effect of increasing the 
gap $\Delta_S$ in the S lead in the absence of any 
direct TS-TS coupling.
The observed supercurrent decrease is consistent with our results in 
Sec.~\ref{sec2c}. This dependence of the current on $
\Delta_S$, with a full suppression for $\Delta_S\to \infty$, confirms that 
a finite TS-TS supercurrent must involve excited states in the S lead.  

\section{Conclusions}\label{sec4}

In this paper, we have studied Josephson transport in a trijunction geometry involving a pair of TS wires and one conventional
$s$-wave (S) superconductor.  One can distinguish different parameter regimes for this TS-S-TS setup. 
In particular, we have studied parallel vs serial configurations, see Fig.~\ref{f1},
where depending on the choice of the bias phases, one either realizes an effective Josephson junction between the S lead and a two-channel TS lead
(parallel configuration), or one has a TS-TS junction with a currentless probe defined by the S lead (serial configuration).  
In the former case, the supercurrent is mainly carried by splitting (creating) Cooper pairs in the S lead into (from) two electrons in different TS wires.
Such processes may become highly effective in the presence of  Majorana
states.  The supercurrent is then maximized (vanishes) for antiparallel (parallel) spin alignment of both TS wires. In the serial configuration,
the picture is basically the other way around since now Josephson transport involves tunneling of quasiparticles between the TS leads.
In the absence of a direct coupling between both TS wires, such processes are only possible via 
excited states in the conventional superconductor.  Interestingly, this configuration also offers a convenient way to stabilize the $4\pi$-periodicity
of the current-phase relation, see Sec.~\ref{sec2c}, and we hope that future experiments will be guided by this observation.

For understanding the physics of such devices, we found it useful to employ a comparatively simple Kitaev chain modeling, see 
Sec.~\ref{sec2}, which allows for analytical progress in several different limits within the boundary Green's function formulation.  
As an alternative approach, we have also extended this formalism to the widely used spinful Majorana wire model of Refs.~\cite{Lutchyn2010,Oreg2010}.
The predictions obtained from the Kitaev model are qualitatively, and in some cases even quantitatively, reproduced from these
 more microscopic calculations.  Most importantly, the spinful wire modeling 
provides access to the spin canting angle $\theta$ between the spin polarization axes in the two TS wires.  We have clarified how this
angle depends on two key parameters of experimental interest, namely on the chemical potential and on the Zeeman field. 
Although not shown explicitly, the present formalism also allows one to take into account finite-length effects in the TS wires \cite{Zazunov2016},
where both Majorana end states in a given TS wire will hybridize.  We expect that the main consequence of this hybridization is 
 a reduction of the critical current due to a decrease in the anomalous Green's function.

To conclude, our results show that the multiterminal Josephson effect in setups containing both conventional and topological
superconductors can provide detailed information about the spin polarization of MBSs.  We hope that our predictions will 
soon be tested experimentally.

\acknowledgments  
We thank R. Aguado, K. Flensberg, T. Jonckheere, T. Martin, and J. Rech  for discussions.  
This work has been supported by the Deutsche Forschungsgemeinschaft (Bonn) 
within Grant No.~EG 96-10/1, 
and by the Spanish MINECO through Grant No.~FIS2014-55486-P
and through the ``Mar\'{\i}a de Maeztu'' Programme for Units of Excellence in R\&D (MDM-2014-0377).

\appendix
\section{On the boundary GF of the spinful model}\label{appa}

Here we provide details on the derivation of the bGF for the spinful nanowire model
in Sec.~\ref{sec3a}.  
For the infinite-wire retarded GF in frequency-momentum space, $\hat G^{(0)}(k,\omega)$, we write the matrix elements
in spin-Nambu space in the form 
\begin{equation}
\hat G^{(0)}_{lm}(k,\omega) = \frac{\tilde{G}^{(0)}_{lm}(k,\omega)}{\det(\omega - \hat H_k)},
\end{equation}
where the indices $l,m=1,2,3,4$ correspond to the entries in the four-spinor $\psi_j$ in Sec.~\ref{sec3a}, and we define $\hat H_k=\hat h+\hat V(k)$, cf.~Eqs.~\eqref{oregmodel} and \eqref{gff0}. 

Using $u_k= 2\alpha\sin(ka)$ 
and $\epsilon_k=2t[1-\cos(ka)]-\mu$,  we obtain the matrix elements
\begin{eqnarray*}
\tilde{G}^{(0)}_{11}(k,\omega) &=& (\omega + \epsilon_k + u_k)[\omega^2-\Delta^2 - (\epsilon_k-u_k)^2]+\\
& +& V_x^2(\epsilon_k - \omega - u_k) ,\\
\tilde{G}^{(0)}_{12}(k,\omega) &= &V_x[\Delta^2 + (\omega + \epsilon_k)^2 - u_k^2 - V_x^2]  ,\\
\tilde{G}^{(0)}_{13}(k,\omega) &= &\Delta[\omega^2 + V_x^2-\Delta^2 - (\epsilon_k-u_k)^2]  ,\\
\tilde{G}^{(0)}_{14}(k,\omega) &= & 2\Delta V_x(\omega + u_k),
\end{eqnarray*} 
\begin{eqnarray*}
\tilde{G}^{(0)}_{21}(k,\omega) &=& V_x[\Delta^2 + (\omega + \epsilon_k)^2 - u_k^2 - V_x^2],  \\
\tilde{G}^{(0)}_{22}(k,\omega) &=& (\omega + \epsilon_k - u_k)[\omega^2-\Delta^2 - (\epsilon_k+u_k)^2] +\\
  &+& V_x^2(\epsilon_k - \omega + u_k) ,\\
\tilde{G}^{(0)}_{23}(k,\omega)& =& 2\Delta V_x(\omega - u_k), \\
\tilde{G}^{(0)}_{24}(k,\omega) &=& \Delta[\omega^2 + V_x^2-\Delta^2 - (\epsilon_k+u_k)^2],  
\end{eqnarray*}
\begin{eqnarray*}
\tilde{G}^{(0)}_{31}(k,\omega)& =& \Delta[\omega^2 + V_x^2-\Delta^2 - (\epsilon_k-u_k)^2] ,  \\
\tilde{G}^{(0)}_{32}(k,\omega)& = &2\Delta V_x(\omega - u_k), \\
\tilde{G}^{(0)}_{33}(k,\omega) &= &(\omega + \epsilon_k- u_k)[-u_k^2-V_x^2 + (\omega - \epsilon_k)^2] +
\\ &+& \Delta^2(\epsilon_k - \omega + u_k), \\
\tilde{G}^{(0)}_{34}(k,\omega)& = &V_x[\Delta^2 + (\omega - \epsilon_k)^2 - u_k^2 - V_x^2] ,
\end{eqnarray*}
\begin{eqnarray*}
\tilde{G}^{(0)}_{41}(k,\omega) &= &2\Delta V_x(\omega + u_k) ,\\
\tilde{G}^{(0)}_{42}(k,\omega)& =& \Delta[\omega^2 + V_x^2-\Delta^2 - (\epsilon_k+u_k)^2] ,  \\ 
\tilde{G}^{(0)}_{43}(k,\omega) &=& V_x[\Delta^2 + (\omega - \epsilon_k)^2 - u_k^2 - V_x^2] ,\\
\tilde{G}^{(0)}_{44}(k,\omega)& =& (\omega + \epsilon_k + u_k)[-u_k^2-V_x^2 + (\omega - \epsilon_k)^2] \\
&-& \Delta^2(-\epsilon_k + \omega + u_k) ,
\end{eqnarray*}
and the determinant
\begin{eqnarray*}
\det(\omega -\hat H_k) &=& \Delta^4 + 2\Delta^2(u_k^2 + \epsilon_k^2 - V_x^2 - \omega^2) + \\
&+& u_k^4 + 2u_k^2(V_x^2 - \omega^2 - \epsilon_k^2) 
\\ 
&+& V_x^4 - 2V_x^2(\omega^2 + \epsilon_k^2) + \omega^4 - 2\epsilon_k^2\omega^2 + \epsilon_k^4.
\end{eqnarray*}


\begin{thebibliography}{99}

\bibitem{Kitaev2001}
A.Yu. Kitaev, Usp. Fiz. Nauk (Suppl) {\bf 171}, 131 (2001).

\bibitem{Alicea2012}
J. Alicea, Rep. Prog. Phys. {\bf 75}, 076501 (2012).

\bibitem{Leijnse2012}
M. Leijnse and K. Flensberg, Semicond. Sci. Techn. {\bf 27}, 124003 (2012).

\bibitem{Beenakker2013}
C.W.J. Beenakker, Annu. Rev. Con. Mat. Phys. {\bf 4}, 113 (2013).

\bibitem{Aasen2016}
D. Aasen \textit{et al.}, Phys. Rev. X {\bf 6}, 031016 (2016).

\bibitem{Mourik2012}
V. Mourik, K. Zuo, S.M. Frolov, S.R. Plissard, E.P.A. Bakkers, and L.P. Kouwenhoven,
Science {\bf 336}, 1003 (2012).

\bibitem{Das2012}
A. Das, Y. Ronen, Y. Most, Y. Oreg, M. Heiblum, and H. Shtrikman, 
Nat. Phys.  {\bf 8}, 887 (2012).

\bibitem{Albrecht2016} 
S.M. Albrecht, A.P. Higginbotham, M. Madsen, F. Kuemmeth, T.S. Jespersen, J. Nyg{\aa}rd, P. Krogstrup, and C.M. Marcus, 
 Nature {\bf 531}, 206 (2016).

\bibitem{Deng2016}
M.T. Deng, S. Vaitiekenas, E.B. Hansen, J. Danon, M. Leijnse, K. Flensberg, J. 
Nyg{\aa}rd, P. Krogstrup, and C.M. Marcus, Science {\bf 354}, 1557 (2016).

\bibitem{Guel2017}
\"O. G\"ul  \textit{et al.}, Nano Lett. {\bf 17}, 2690 (2017).

\bibitem{Albrecht2017}
S.M. Albrecht, E.B. Hansen, A.P. Higginbotham, F. Kuemmeth, T.S. Jespersen, J. Nyg{\aa}rd, P. Krogstrup, J. Danon, K. Flensberg and C.M. Marcus, 
Phys. Rev. Lett. {\bf 118}, 137701 (2017).

\bibitem{Zhang2017}
H. Zhang \textit{et al.},  arXiv:1603.04069.

\bibitem{Suominen2017}
H.J. Suominen, M. Kjaergaard, A.R. Hamilton, J. Shabani,
C.J. Palmstr{\o}m, C.M. Marcus, and F. Nichele, arXiv:1703.03699.

\bibitem{Yazdani2014}
S. Nadj-Perge, I.K. Drozdov, J. Li, H. Chen, S. Jeon, J. Seo, A.H. MacDonald, B.A. Bernevig, and A. Yazdani, Science {\bf 346}, 602 (2014).

\bibitem{Franke2015}
M. Ruby, F. Pientka, Y. Peng, F. von Oppen, B.W. Heinrich, and K.J. Franke,   
Phys. Rev. Lett. {\bf 115}, 197204 (2015).
 
 \bibitem{Sun2016}
 H.H. Sun \textit{et al.}, Phys. Rev. Lett. {\bf 116}, 257003 (2016).
 
 \bibitem{Feldman2017}
 B.E. Feldman, M.T. Randeria, J. Li, S. Jeon, Y. Xie, Z. Wang, I.K. Drozdov,
 B. Andrei Bernevig, and A. Yazdani, Nat. Phys. {\bf 13}, 286 (2017).
 
 \bibitem{Deacon2017}
R.S. Deacon \textit{et al.}, Phys. Rev. X {\bf 7}, 021011 (2017).

\bibitem{Gazi2017}
S. Gazibegovich \textit{et al.}, arXiv:1705.01480.

\bibitem{Zazunov2016}
A. Zazunov, R. Egger, and A. Levy Yeyati, Phys. Rev. B {\bf 94}, 014502 (2016).

\bibitem{Peng2017}
Y. Peng, Y. Bao, and F. von Oppen, Phys. Rev. B {\bf 95}, 235143 (2017).

\bibitem{Cuevas1996}
 J.C. Cuevas, A. Mart{\'i}n-Rodero, and A. Levy Yeyati, 
 Phys. Rev. B {\bf 54}, 7366 (1996).

\bibitem{Nazarov2009}
Yu.V. Nazarov and Ya.M. Blanter, \textit{Quantum Transport: Introduction to Nanoscience} (Cambridge University Press, Cambridge, UK, 2010).

\bibitem{Jonckheere2017}
T. Jonckheere, J. Rech, A. Zazunov, R. Egger, and T. Martin, Phys. Rev. B 
{\bf 95}, 054514 (2017).

\bibitem{Lutchyn2010}
R.M. Lutchyn, J.D. Sau, and S. Das Sarma, Phys. Rev. Lett. {\bf 105}, 077001 (2010).

\bibitem{Oreg2010}
Y. Oreg, G. Refael, and F. von Oppen, Phys. Rev. Lett. {\bf 105}, 177002 (2010).

\bibitem{Flensberg2011}
M. Leijnse and K. Flensberg, Phys. Rev. Lett. {\bf 107}, 210502 (2011).

\bibitem{Sticlet2012} D. Sticlet, C. Bena, and P. Simon, Phys. Rev. Lett. {\bf 108}, 096802 (2012).

\bibitem{Prada2012}
E. Prada, P. San-Jose, and R. Aguado, Phys. Rev. B {\bf 86}, 180503(R) (2012).

\bibitem{Rainis2013}
D. Rainis, L. Trifunovic, J. Klinovaja, and D. Loss, 
Phys. Rev. B {\bf 87}, 024515 (2013).

\bibitem{Jiang2013}
L. Jiang, D. Pekker, J. Alicea, G. Refael, Y. Oreg, A. Brataas, and F. von Oppen,
Phys. Rev. B {\bf 87}, 075438 (2013).

\bibitem{He2014}
J.J. He, T.K. Ng, P.A. Lee, and K.T. Law, Phys. Rev. Lett. {\bf 112}, 037001 (2014).

\bibitem{Prada2017} E. Prada, R. Aguado, and P. San-Jose, arXiv:1702.02525.

\bibitem{Hoffman2017} S. Hoffman, D. Chevallier, D. Loss, and J. Klinovaja, arXiv:1705.03002.

\bibitem{Zazunov2012} 
A. Zazunov and R. Egger, 
 Phys. Rev.  B {\bf 85}, 104514 (2012).

\bibitem{Peng2015}
Y. Peng, F. Pientka, Y. Vinkler-Aviv, L.I. Glazman, and F. von Oppen, 
Phys. Rev. Lett. {\bf 115}, 266804 (2015).

\bibitem{Ioselevich2016}
P.A. Ioselevich, P.M. Ostrovsky, and M.V. Feigelman, 
  Phys. Rev. B {\bf 93}, 125435 (2016).

\bibitem{Sharma2016}
G. Sharma and S. Tewari, Phys. Rev. B {\bf 93}, 195161 (2016).

\bibitem{Setiawan2017a}
F. Setiawan, W.S. Cole, J.D. Sau, and S. Das Sarma,
Phys. Rev. B {\bf 95}, 174515 (2017).

\bibitem{Setiawan2017b}
F. Setiawan, W.S. Cole, J.D. Sau, and S. Das Sarma,
Phys. Rev. B {\bf 95}, 020501(R) (2017).

\bibitem{Nilsson2008}
J. Nilsson,  A.R. Akhmerov, and C.W.J. Beenakker,  
Phys. Rev. Lett. {\bf 101}, 120403 (2008).

\bibitem{Alicea2011}
J. Alicea, Y. Oreg, G. Refael, F. von Oppen, and M.P.A. Fisher,
Nat. Phys. {\bf 7}, 412 (2011).

\bibitem{Jiang2011}
L. Jiang, D. Pekker, J. Alicea, G. Refael, Y. Oreg, and F. von Oppen, 
Phys. Rev. Lett. {\bf 107}, 236401 (2011).

\bibitem{Riwar2016}
R.P. Riwar, M. Houzet, J.S. Meyer, and Y. Nazarov,
Nat. Comm. {\bf 7}, 11167 (2016).

\bibitem{foot1}
Our model describes a spin-polarized $p$-wave TS wire as well as a helical TS wire, where right- and left-movers have opposite spin.
These two case imply $\theta_j= 0$ and $\theta_j = \pi/2$, respectively.
For pointlike tunneling contacts, see Eq.~\eqref{Ht}, left- and right-movers enter with equal amplitude, and both
two cases are indistinguishable due to the SU(2) spin symmetry of the S lead.

\bibitem{Schapers2010}
S. Est{\'e}vez Hern{\'a}ndez, M. Akabori, K. Sladek, Ch. Volk, S. Alagha, H. Hardtdegen, M.G. Pala, N. Demarina, D. Gr\"utzmacher, and Th. Sch\"apers,
Phys. Rev. B {\bf 82}, 235303 (2010).

\end{thebibliography}
\end{document}